\documentclass[preprintnumbers,unsortedaddress,superscriptaddress,notitlepage]{revtex4}
\usepackage{amsfonts}
\usepackage{amsmath}
\usepackage{hyperref}
\usepackage{amssymb}
\usepackage[english]{babel}
\usepackage{graphicx}
\usepackage{epsfig}
\usepackage{bm}
\usepackage{longtable}
\usepackage{verbatim}
\usepackage{longtable}

\newcommand{\mathsym}[1]{{}}

\newcommand{\emp}{\begin{equation}}
\newcommand{\fin}{\end{equation}}
\newcommand{\empn}{\begin{equation*}}
\newcommand{\finn}{\end{equation*}}

\newcommand{\bea}{\begin{eqnarray}}
\newcommand{\eea}{\end{eqnarray}}
\newcommand{\eger}{\begin{gather}}
\newcommand{\fger}{\end{gather}}
\newcommand{\egn}{\begin{gather*}}
\newcommand{\fgn}{\end{gather*}}
\newcommand{\bit}{\begin{itemize}}
\newcommand{\eit}{\end{itemize}}

\input{tcilatex}
\begin{document}

\title{An SU(5) grand unified model with discrete flavour symmetries.}
\author{A. E. C\'arcamo Hern\'andez}
\email{antonio.carcamo@usm.cl}
\affiliation{{\small Universidad T\'ecnica Federico Santa Mar\'{\i}a and Centro Cient%
\'{\i}fico-Tecnol\'ogico de Valpara\'{\i}so}\\
Casilla 110-V, Valpara\'{\i}so, Chile}
\author{Sergey Kovalenko}
\email{sergey.kovalenko@usm.cl}
\affiliation{{\small Universidad T\'ecnica Federico Santa Mar\'{\i}a and Centro Cient%
\'{\i}fico-Tecnol\'ogico de Valpara\'{\i}so}\\
Casilla 110-V, Valpara\'{\i}so, Chile}
\author{Iv\'an Schmidt}
\email{ivan.schmidt@usm.cl}
\affiliation{{\small Universidad T\'ecnica Federico Santa Mar\'{\i}a and Centro Cient%
\'{\i}fico-Tecnol\'ogico de Valpara\'{\i}so}\\
Casilla 110-V, Valpara\'{\i}so, Chile}
\date{\today }

\begin{abstract}
We propose a model based on the $SU(5)$ grand unification with an extra $%
Z_{2}\otimes Z_{2}^{\prime }\otimes Z_{2}^{\prime \prime }\otimes
Z_{4}\otimes Z_{12}$ flavor symmetry, which successfully describes the
observed SM fermion mass and mixing pattern. 
The observed quark mass and mixing pattern is caused by the $Z_{4}$ and $%
Z_{12}$ symmetries, which are broken at very high scale by the $SU(5)$
scalar singlets $\sigma $ and $\chi $, charged respectively under these
symmetries and which acquire VEVs at the GUT scale. The light neutrino
masses are generated via a type I seesaw mechanism with three heavy Majorana
neutrinos. The model has in total 17 effective free parameters, from which 2
are fixed and 15 are fitted to reproduce the experimental values of the 18
physical parameters in the quark and lepton sectors. The model predictions
for both quark and lepton sectors are in excellent agreement with the
experimental data. 
\end{abstract}

\maketitle


\section{Introduction}

In spite of the Standard Model (SM) great success in 
describing electroweak phenomena, recently confirmed with the LHC discovery
of a $126$ GeV Higgs boson \cite{LHC-H-discovery}, it has many open
questions \cite{SM,PDG}. 
Among the most pressing are the smallness of neutrino masses, 
the fermion mass and mixing hierarchy, and the existence of the three
generations of fermions. 
The existing pattern of fermion masses goes over a range of five orders of
magnitude in the quark sector and a much wider range when neutrinos are
included. While the mixing angles in the quark sector are very small, in the
lepton sector two of the mixing angles are large, and one mixing angle is
small. This suggests a different kind of New Physics for the neutrino sector
from the one present in the quark mass and mixing pattern. Experiments with
solar, atmospheric and reactor neutrinos have brought clear evidence of
neutrino oscillations from the measured non vanishing neutrino mass squared
splittings. This brings compelling and indubitable evidence that at least
two of the neutrinos have non vanishing masses, much smaller, by many orders
of magnitude, than the SM charged fermion masses, and that the three
neutrino flavors mix. 

\quad The global fits of the available data from neutrino oscillation
experiments Daya Bay \cite{An:2012eh}, T2K \cite{Abe:2011sj}, MINOS \cite%
{Adamson:2011qu}, Double CHOOZ \cite{Abe:2011fz} and RENO \cite{Ahn:2012nd},
constrain the neutrino mass squared splittings and mixing parameters, as
shown in Tables \ref{NH} and \ref{IH} (based on Ref. \cite{Forero:2014bxa})
for the normal (NH) and inverted (IH) hierarchies of the neutrino mass
spectrum. These facts might suggest that the tiny neutrino masses can be
related to a scale of New Physics that, in general, is not related to the
scale of Electroweak Symmetry Breaking (EWSB) $v=246$ GeV. 
\begin{table}[tbh]
\begin{tabular}{|c|c|c|c|c|c|}
\hline
Parameter & $\Delta m_{21}^{2}$($10^{-5}$eV$^2$) & $\Delta m_{31}^{2}$($%
10^{-3}$eV$^2$) & $\left( \sin ^{2}\theta _{12}\right) _{\exp }$ & $\left(
\sin ^{2}\theta _{23}\right) _{\exp }$ & $\left( \sin ^{2}\theta
_{13}\right) _{\exp }$ \\ \hline
Best fit & $7.60$ & $2.48$ & $0.323$ & $0.567$ & $0.0234$ \\ \hline
$1\sigma $ range & $7.42-7.79$ & $2.41-2.53$ & $0.307-0.339$ & $0.439-0.599$
& $0.0214-0.0254$ \\ \hline
$2\sigma $ range & $7.26-7.99$ & $2.35-2.59$ & $0.292-0.357$ & $0.413-0.623$
& $0.0195-0.0274$ \\ \hline
$3\sigma $ range & $7.11-8.11$ & $2.30-2.65$ & $0.278-0.375$ & $0.392-0.643$
& $0.0183-0.0297$ \\ \hline
\end{tabular}%
\caption{Range for experimental values of neutrino mass squared splittings
and leptonic mixing parameters, taken from Ref. \protect\cite{Forero:2014bxa}%
, for the case of normal hierarchy.}
\label{NH}
\end{table}
\begin{table}[tbh]
\begin{tabular}{|c|c|c|c|c|c|}
\hline
Parameter & $\Delta m_{21}^{2}$($10^{-5}$eV$^{2}$) & $\Delta m_{13}^{2}$($%
10^{-3}$eV$^{2}$) & $\left( \sin ^{2}\theta _{12}\right) _{\exp }$ & $\left(
\sin ^{2}\theta _{23}\right) _{\exp }$ & $\left( \sin ^{2}\theta
_{13}\right) _{\exp }$ \\ \hline
Best fit & $7.60$ & $2.38$ & $0.323$ & $0.573$ & $0.0240$ \\ \hline
$1\sigma $ range & $7.42-7.79$ & $2.32-2.43$ & $0.307-0.339$ & $0.530-0.598$
& $0.0221-0.0259$ \\ \hline
$2\sigma $ range & $7.26-7.99$ & $2.26-2.48$ & $0.292-0.357$ & $0.432-0.621$
& $0.0202-0.0278$ \\ \hline
$3\sigma $ range & $7.11-8.11$ & $2.20-2.54$ & $0.278-0.375$ & $0.403-0.640$
& $0.0183-0.0297$ \\ \hline
\end{tabular}%
\caption{Range for experimental values of neutrino mass squared splittings
and leptonic mixing parameters, taken from Ref. \protect\cite{Forero:2014bxa}%
, for the case of inverted hierarchy.}
\label{IH}
\end{table}

\quad The flavour puzzle of the SM indicates that New Physics has to be
advocated to explain the prevailing patterm of fermion masses and mixings.
To tackle the limitations of the SM, various extensions of the SM including
larger scalar and/or fermion sector as well as extended gauge group with
additional flavor symmetries, have been proposed in the literature (for a
reviews see, e.g., Refs. \cite%
{Fritzsch:1999ee,Altarelli:2002hx,Altarelli:2010gt,Ishimori:2010au,FlavorSymmRev}%
). Another approach to describe the fermion mass and mixing pattern consists
in postulating particular mass matrix textures (see Ref \cite{Textures} for
some works considering textures). 
Concerning models with an extended gauge symmetry, Grand unified theories
(GUTs) endowed with flavor symmetries may provide an unified description for
the mass and mixing pattern of leptons and quarks. 
This is motivated by the fact that leptons and quarks belong to the same
multiplets of the GUT group, allowing to relate their masses and mixings 
\cite{Marzocca:2011dh,Antusch:2013kna}. This framework is also very useful
for explaining the smallness of neutrino masses through the simplest type I
seesaw mechanism, where the new heavy Majorana neutrinos have masses at the
GUT scale. 
Various GUT models with flavor symmetries have been proposed in the
literature 
\cite%
{Chen:2013wba,King:2012in,Meloni:2011fx,BhupalDev:2012nm,Babu:2009nn,Babu:2011mv,Gomez-Izquierdo:2013uaa,Antusch:2010es,Hagedorn:2010th,Ishimori:2008fi,Patel:2010hr,Chen:2007}%
. 
For a general review see for example \cite{King:2013eh,Chen:2003zv}.

\quad Recently we proposed a model based on the $SU(5)$ grand unification
with an extra $A_{4}\otimes Z_{2}\otimes Z_{2}^{\prime }\otimes
Z_{2}^{\prime \prime }\otimes U\left( 1\right) _{f}$ flavor symmetry \cite%
{Campos:2014lla}, which successfully accounts for the SM fermion mass and
mixing pattern. That model involves a horizontal symmetry $U_{f}(1)$, which
provides an explanation for the prevailing pattern of charged fermion masses
and quark mixing matrix elements, by means of generalized Froggatt-Nielsen
mechanism \cite{Wang:2011ub}. In that model, the light neutrino masses are
generated via a radiative seesaw mechanism, with a single heavy Majorana
neutrino and neutral scalars running in the loops. Nevertheless, that model
has a non minimal scalar sector, and at low energies reduces to an eight
Higgs doublet model (8HDM) with a light scalar octet, thus making it not
predictive in the scalar sector. Furthermore, in that model the obtained
values for the observables in the quark and lepton sector are in good
agreement with the experimental data, with the exception of the up and charm
quark masses.

\quad It is interesting to find an alternative and better explanation for
the SM fermion mass and mixing hierarchy, by formulating a $SU(5)$ grand
unification with less scalar content than our previous model of Ref. \cite%
{Campos:2014lla}. To this end, we propose an alternative and improved
version of the $SU(5)$ GUT model with an additional flavor symmetry group 
\mbox{$Z_{2}\otimes Z_{2}^{\prime }\otimes Z_{2}^{\prime \prime }\otimes
Z_{4}\otimes Z_{12}$}, which is consistent with the current data on fermion
masses and mixings. The particular role of each discrete symmetry is
explained in the following. The $Z_{2}$ symmetry separates the scalars
participating in the Yukawa interactions for charged leptons and down type
quarks from those ones participating in the Yukawa interactions for up type
quarks. This results in a separation of the up type quark sector from the
down type quark and charged lepton sector, thus reducing the number of model
parameters. The $Z_{2}^{\prime }$ symmetry determines the allowed entries of
the mass matrices for down type quarks, charged leptons and neutrinos. The $%
Z_{2}^{\prime \prime }$ symmetry separates the heavy right handed Majorana
neutrinos from the remaining fermionic fields. The $Z_{4}$ symmetry is
crucial for explaining the smallness of the down quark and electron masses.
Without this symmetry they both would be larger than their corresponding
experimental values by about two orders of magnitude, unless one sets their
Yukawa couplings unnaturally small. 
The $Z_{12}$ symmetry induces most of the charged fermion mass and quark
mixing hierarchy. Let us recall that due to the properties of the $Z_{N}$
groups, it follows that $Z_{12}$ is the lowest cyclic symmetry that allows
to build a ten dimensional up type quark Yukawa term, crucial to get the
required $\lambda ^{6}$ supression in the 11 entry of the up type quark mass
matrix, where $\lambda =0.225$ is one of the Wolfenstein parameters. This
symmetry is essential in order to get the observed pattern of charged
fermion masses and quark mixings. 
The scalar sector of our model includes the following $SU\left( 5\right) $
representations: one $\mathbf{24}$, one $\mathbf{45}$'s, three $\mathbf{5}$%
's and four $\mathbf{1}$'s. The four $SU\left( 5\right) $ scalar singlets
and the scalar in the $\mathbf{24}$ irrep of $SU\left( 5\right) $ acquire
vacuum expectation values (VEVs) at the GUT scale. 
The particular role of each additional scalar field and the corresponding
particle assignments under the symmetry group of the model are explained in
detail in Sec. \ref{model}. In the present model the fermion sector is
extended by introducing three heavy right handed Majorana neutrinos, which
are singlets under the SM group. These heavy right handed Majorana
neutrinos, which acquire masses at the GUT scale due to their interactions
with the $SU(5)$ singlet scalar fields, allow us to generate small active
neutrino masses through type I seesaw mechanism. In this framework, the
active neutrinos acquire small masses scaled by the inverse of the large
Majorana neutrino masses, thus providing a natural explanation for the
smallness of neutrino masses. 

\quad Our model at low energies corresponds to a four Higgs doublet model
(4HDM), with a light scalar color octet and is more minimal than several
models proposed in the literature, such as \cite%
{Antusch:2010es,Babu:2011mv,Gomez-Izquierdo:2013uaa,Campos:2014lla}. 
Our model successfully describes a realistic pattern of the SM fermion
masses and mixings. The model has 16 free effective parameters, from which 2
are fixed and 14 are fitted to 
reproduce the experimental values of 18 observables, i.e., 9 charged fermion
masses, 2 neutrino mass squared splittings, 3 lepton mixing parameters, 3
quark mixing angles and 1 CP violating phase of the CKM quark mixing matrix.
All obtained physical parameters in the quark and lepton sector are in
excellent agreement with the experimental data.

\quad The paper is organized as follows. In Sec. \ref{model} we explain
theoretical aspects of the proposed model. 
In Sec. \ref{quarkmassesandmixing} we focus on the discussion of quark
masses and mixing and give our corresponding results. Our results regarding
lepton masses and mixing, followed by a numerical analysis, are presented in
Sec. \ref{leptonmassmix}. Conclusions are given in Sec. \ref{conclusions}. 

\section{The Model}

\label{model} As is well known, the minimal $SU\left( 5\right) $ GUT \cite%
{Georgi:1974sy} with fermions in $\mathbf{\bar{5}}+\mathbf{10}$ and the
scalars in $\mathbf{5}+\mathbf{24}$ representations of $SU\left( 5\right) $,
has several drawbacks. In particular, it predicts wrong relations between
the down-type quark and charged lepton masses, short proton life-time, and
the unification of gauge couplings is not consistent with the values of $%
\alpha _{S}$, $\sin \theta _{W}$ and $\alpha _{em}$ at the $M_{Z}$ scale.
The minimal model does not account for non vanishing neutrino masses,
contradicting neutrino oscillation experiments. %
Addresing some of these problems requires an extension of the model scalar
sector, by including, in particular, a scalar $\mathbf{45}$ representation
of $SU(5)$ \cite%
{Georgi:1979df,Frampton:1979,Ellis:1979,Nandi:1980sd,Frampton:1980,Langacker:1980js,Kalyniak:1982pt,Giveon:1991,Dorsner:2007fy,Dorsner:2006dj,FileviezPerez:2007nh,Perez:2008ry,Khalil:2013ixa}%
. 
However, the next-to-minimal $SU\left( 5\right) $ GUT model is
unsatisfactory in describing the fermion mass and mixing pattern, due to the
unexplained hierarchy among the large number of Yukawa couplings in the
model. To address that problem, we recently proposed a model based on the $%
SU(5)$ grand unification with an extra $A_{4}\otimes Z_{2}\otimes
Z_{2}^{\prime }\otimes Z_{2}^{\prime \prime }\otimes U\left( 1\right) _{f}$
flavor symmetry, which successfully accounts for the SM fermion mass and
mixing pattern. In that model the fermion mass hierarchy is explained by a
spontaneously broken group $U(1)_{f}$ with a special $U(1)_{f}$ charge
assignment to the fields participating in the Yukawa terms. However, that
model has a non minimal scalar sector, and at low energies reduces to an
eight Higgs doublet model with a light scalar octet, which is not predictive
in the scalar sector. Therefore it would be desirable to explain the SM
fermion mass and mixing hierarchy by formulating a $SU(5)$ grand unification
model with a more minimal scalar content than our previous model of Ref. 
\cite{Campos:2014lla}. To this end, we consider a multi-Higgs extension of
the next-to-minimal $SU\left( 5\right) $ GUT, with the full symmetry $%
\mathcal{G}$ experiencing a two-step spontaneous breaking: 
\begin{eqnarray}
&&\mathcal{G}=SU\left( 5\right) \otimes Z_{2}\otimes Z_{2}^{\prime }\otimes
Z_{2}^{\prime \prime }\otimes Z_{4}\otimes Z_{12}  \label{Group} \\
&&\hspace{35mm}\Downarrow \Lambda _{GUT}  \notag \\[3mm]
&&\hspace{15mm}SU\left( 3\right) _{C}\otimes SU\left( 2\right) _{L}\otimes
U\left( 1\right) _{Y}\otimes Z_{2}\otimes Z_{2}^{\prime }  \notag \\[3mm]
&&\hspace{35mm}\Downarrow \Lambda _{EW}  \notag \\[3mm]
&&\hspace{23mm}SU\left( 3\right) _{C}\otimes U\left( 1\right) _{em}  \notag
\end{eqnarray}%
\quad A relevant difference of this Grand Unified Model with our previous $%
SU(5)$ model with $A_{4}$ flavour symmetry is that the former does not
involve neither the global $U\left( 1\right) $ symmetry crucial to trigger
the generalized Froggatt-Nielsen mechanism nor the $A_{4}$ flavour symmetry
while the later does. Our current $SU(5)$ GUT model involves instead a set
of $Z_{2}$, $Z_{2}^{\prime }$, $Z_{2}^{\prime \prime }$, $Z_{4}$ and $Z_{12}$
discrete symmetries. Furthermore, our $SU(5)$ GUT model with $A_{4}$ flavour
symmetry of Ref. \cite{Campos:2014lla} includes the $Z_{2}$, $Z_{2}^{\prime }
$ and $Z_{2}^{\prime \prime }$ discrete symmetries, with a unbroken $Z_{2}$
symmetry, while in our current $SU(5)\otimes Z_{2}\otimes Z_{2}^{\prime
}\otimes Z_{2}^{\prime \prime }\otimes Z_{4}\otimes Z_{12}$ GUT model, the $%
Z_{2}$ symmetry is broken as well as the remaining discrete symmetries.
While the linear combinations of $U\left( 1\right) $ charges of the fields
participating in the Yukawa terms are put by hand in our $SU(5)$ GUT model
with $A_{4}$ flavour symmetry in order to generate specific fermion mass
matrix textures, in our current $SU(5)$ GUT model the hierarchy among the
fermion masses naturally arises from the $Z_{4}\otimes Z_{12}$ charge
assignments of the fermion and scalar fields. In particular, the $Z_{12}$
symmetry together with the $Z_{4}$ symmetries will be crucial for explaining
the smallness of the first generation charged fermions. Besides that, the $%
Z_{4}\otimes Z_{12}$ symmetry will shape the hierarchical structure of the
up and down type quark mass matrices necessary to get a realistic pattern of
quark masses and mixings. The $Z_{2}$ symmetry will separate the up type
quark sector from the down type quark and charged lepton sector resulting in
a reduction of the number of model parameters. Let us recall that due to the
properties of the $Z_{N}$ groups, it follows that $Z_{12}$ is the lowest
cyclic symmetry that allows to build a ten dimensional up type quark Yukawa
term, crucial to get the required $\lambda ^{6}$ supression in the 11 entry
of the up type quark mass matrix, where $\lambda =0.225$ is one of the
Wolfenstein parameters. The $Z_{2}^{\prime }$ symmetry will determine the
allowed entries of the mass matrices for down type quarks and charged
leptons. The $Z_{2}^{\prime \prime }$ symmetry separates the heavy right
handed Majorana neutrinos from the remaining fermionic fields. The $Z_{4}$
symmetry is crucial for explaining the smallness of the down quark and
electron masses. Without this symmetry they both would be larger by about
two orders of magnitude than their corresponding experimental values, unless
one sets the corresponding Yukawa couplings unnaturally small. Furthermore,
while the CKM matrix is fitted to the experimental data in the $SU(5)$ GUT
model with $A_{4}$ flavour symmetry, our current $SU(5)$ GUT model predicts
a specific hierarchical structure for the CKM matrix, consistent with the
experimental data. All the aforementioned features make our current model an
important improvement of our previous $SU(5)$ GUT model with $A_{4}$ flavour
symmetry of Ref. \cite{Campos:2014lla}. In the present model the fermion
sector is extended by introducing three heavy right handed Majorana
neutrinos, which are singlets under the SM group. The fermion assignments
under the group $\mathcal{G}=SU(5)\otimes Z_{2}\otimes Z_{2}^{\prime
}\otimes Z_{2}^{\prime \prime }\otimes Z_{4}\otimes Z_{12}$ are: 
\begin{eqnarray}
\psi ^{i\left( 1\right) } &\sim &\left( \overline{\mathbf{5}}%
,-1,-1,1,-1,-i\right) ,\hspace{1.5cm}\psi ^{i\left( 2\right) }\sim \left( 
\overline{\mathbf{5}},-1,1,1,1,-i\right) ,\hspace{1.5cm}\psi ^{i\left(
3\right) }\sim \left( \overline{\mathbf{5}},-1,1,1,1,-i\right) ,  \notag \\
\Psi _{ij}^{\left( 1\right) } &\sim &\left( \mathbf{10,}1,1,1,1,i\right) ,%
\hspace{0.5cm}\hspace{0.5cm}\Psi _{ij}^{\left( 2\right) }\sim \left( \mathbf{%
10,}1,1,1,1,\omega \right) ,\hspace{0.5cm}\hspace{0.5cm}\Psi _{ij}^{\left(
3\right) }\sim \left( \mathbf{10,}1,1,1,1,1\right) ,\hspace{0.5cm}%
i,j=1,2,3,4,5.  \notag \\
N_{R}^{\left( 1\right) } &\sim &\left( \mathbf{1,-}1,-1,-1,1,1\right) ,%
\hspace{0.5cm}\hspace{0.5cm}\hspace{0.5cm}N_{R}^{\left( 2\right) }\sim
\left( \mathbf{1,-}1,1,-1,1,1\right) ,\hspace{0.5cm}\hspace{0.5cm}\hspace{%
0.5cm}N_{R}^{\left( 3\right) }\sim \left( \mathbf{1,-}1,1,-1,1,-1\right) 
\label{fermions}
\end{eqnarray}%
where $\omega =e^{\frac{2\pi i}{3}}$.

\quad More explicitly, we accomodate the fermions as follows: 
\begin{equation}
\Psi _{ij}^{\left( f\right) }=\frac{1}{\sqrt{2}}\left( 
\begin{array}{ccccc}
0 & u_{3}^{\left( f\right) c} & -u_{2}^{\left( f\right) c} & -u_{1}^{\left(
f\right) } & -d_{1}^{\left( f\right) } \\ 
-u_{3}^{\left( f\right) c} & 0 & u_{1}^{\left( f\right) c} & -u_{2}^{\left(
f\right) } & -d_{2}^{\left( f\right) } \\ 
u_{2}^{\left( f\right) c} & -u_{1}^{\left( f\right) c} & 0 & -u_{3}^{\left(
f\right) } & -d_{3}^{\left( f\right) } \\ 
u_{1}^{\left( f\right) } & u_{2}^{\left( f\right) } & u_{3}^{\left( f\right)
} & 0 & -l^{\left( f\right) c} \\ 
d_{1}^{\left( f\right) } & d_{2}^{\left( f\right) } & d_{3}^{\left( f\right)
} & l^{\left( f\right) c} & 0%
\end{array}%
\right) _{L},\hspace{1.5cm}f=1,2,3\hspace{1.5cm}i,j=1,2,3,4,5.
\end{equation}%
\begin{equation}
\psi ^{i\left( f\right) }=\left( d_{1}^{\left( f\right) c},d_{2}^{\left(
f\right) c},d_{3}^{\left( f\right) c},l^{\left( f\right) },-\nu _{f}\right)
_{L}.
\end{equation}%
Where the subindices denote the different quark colors whereas the
superscript $f$ labels the fermion families.

\quad The scalar sector is composed of the following $SU\left( 5\right) $
representations: one $\mathbf{24}$, one $\mathbf{45}$'s, three $\mathbf{5}$%
's and four $\mathbf{1}$'s. Thus, the scalar fields of our model have the
following $\mathcal{G}$ assignments: 
\begin{eqnarray}
\chi &\sim &\left( \mathbf{1,}1,1,1,1,\omega ^{\frac{1}{4}}\right) ,\hspace{%
1.5cm}\sigma \sim \left( \mathbf{1,}1,1,1,1,i,1\right) ,\hspace{1.5cm}\eta
\sim \left( \mathbf{1,}1,1,-1,-1,1\right)  \notag \\
\zeta &\sim &\left( \mathbf{1,}1,1,-1,1,1\right) ,\hspace{1cm}H_{i}^{\left(
1\right) }\sim \left( \mathbf{5,}1,1,1,1,1\right) ,\hspace{1.5cm}%
H_{i}^{\left( 2\right) }\sim \left( \mathbf{5,}1,-1,1,1,1\right) ,  \notag \\
H_{i}^{\left( 3\right) } &\sim &\left( \mathbf{5,-}1,1,1,1,1\right) ,\hspace{%
1cm}\Sigma _{j}^{i}\sim \left( \mathbf{24,}1,1,1,1,1\right) ,\hspace{1.5cm}%
\Phi _{jk}^{i}\sim \left( \mathbf{45,-}1,1,1,1,1\right) .  \label{scalars}
\end{eqnarray}

Note that the aforementioned scalar content of our model is much more
minimal than the corresponding to our previous model of $SU(5)$ GUT model
with $A_{4}$ flavour symmetry of Ref. \cite{Campos:2014lla}, which includes
one $\mathbf{24}$, one $\mathbf{45}$, seven $\mathbf{5}$'s and six $\mathbf{1%
}$'s irreps of $SU\left( 5\right) $. As previously mentioned, the scalar
field $\Sigma $ gets a vacuum expectation value (VEV) at the GUT scale $%
\Lambda _{GUT}=10^{16}$ GeV and triggers the first step of symmetry breaking
in Eq. (\ref{Group}). This first step is also induced by the scalars $\chi $
, $\sigma $, $\eta $ and $\zeta $, which get VEVs at the GUT scale. The
second step of symmetry breaking, is caused by the scalars $H_{i}^{\left(
f\right) }$ ($f=1,2,3$) and $\Phi _{jk}^{i}$ acquiring VEVs at the scale of
electroweak symmetry breaking.

\quad Our current model is based on the following assumptions:

\begin{enumerate}
\item The symmetry of the $SU\left( 5\right) $ GUT Model is extended to
include the discrete symmetries $Z_{2}$, $Z_{2}^{\prime }$, $Z_{2}^{\prime
\prime }$, $Z_{4}$ and $Z_{12}$. The $Z_{2}$, $Z_{2}^{\prime }$\ and $%
Z_{2}^{\prime \prime }$, $Z_{4}$, $Z_{12}$ discrete symmetries are broken at
the Electroweak and GUT scales, respectively. 

\item The scalar sector includes the following $SU\left( 5\right) $
representations: one $\mathbf{24}$, one $\mathbf{45}$'s, three $\mathbf{5}$%
's and four $\mathbf{1}$'s. The four $SU\left( 5\right) $ scalar singlets
and the scalar in the $\mathbf{24}$ irrep of $SU\left( 5\right) $ acquire
VEVs at the GUT scale. The scalar field in the $\mathbf{24}$ irrep of $%
SU\left( 5\right) $ is needed to trigger the first step of symmetry breaking
in Eq. (\ref{Group}), which is also induded by the $\mathbf{1}$'s irreps of $%
SU\left( 5\right) $. The remaining scalars acquire VEVs at the electroweak
scale and induce the second step of symmetry breaking.\ As previously
mentioned, having scalar fields in the $\mathbf{45}$ representation of $%
SU\left( 5\right) $ is crucial to get the correct mass relations of
down-type quarks and charged leptons.

\item The $Z_{2}$ symmetry separates the scalars in the $\mathbf{5}$ and $%
\mathbf{45}$ irreps of $SU\left( 5\right) $ participating in the Yukawa
interactions for charged leptons and down type quarks from those ones
participating in the Yukawa interactions for up type quarks. This implies
the $SU\left( 5\right) $ scalar multiplets contributing to the masses of the
down-type quarks and charged leptons are different from those ones providing
masses to the up-type quarks. The fermions belonging to the $\mathbf{10}$
irrep of $SU\left( 5\right) $ are $Z_{2}$ even while those ones embedding in
the $\overline{\mathbf{5}}$ irrep of $SU\left( 5\right) $ are $Z_{2}$ odd.
The $\mathbf{45}$ and one of the $\mathbf{5}$'s scalars are $Z_{2}$ odd and
thus they participate in the Yukawa interactions for charged leptons and
down type quarks. The remaining two $\mathbf{5}$'s, which are $Z_{2}$ even
participate in the Yukawa interactions for up type quarks. The three scalars 
$SU\left( 5\right) $ singlets are $Z_{2}$ even.

\item The $Z_{2}^{\prime }$ symmetry determines the allowed entries of the
mass matrices for down type quarks, charged leptons and neutrinos. The $%
Z_{2}^{\prime }$ symmetry separates the $Z_{2}^{\prime }$ odd fermionic $%
\overline{\mathbf{5}}^{\left( 1\right) }$\ irrep of $SU\left( 5\right) $
belonging to the first family from the remaining fermionic $\overline{%
\mathbf{5}}^{\left( 2\right) }$ and $\overline{\mathbf{5}}^{\left( 3\right)
} $ irreps of $SU\left( 5\right) $, neutral under this symmetry.
Furthermore, the $Z_{2}^{\prime }$ symmetry separates the first generation
right handed heavy Majorana neutrino $N_{R}^{\left( 1\right) }$, neutral
under this symmetry, from the second and third generation ones, i.e., $%
N_{R}^{\left( 2\right) }$ and $N_{R}^{\left( 3\right) }$, charged under the $%
Z_{2}^{\prime }$ symmetry. \ Thus, the $Z_{2}^{\prime }$ symmetry forbidds
mixings of the first generation right handed heavy Majorana neutrino $%
N_{R}^{\left( 1\right) }$ with the second and third generation ones, i.e., $%
N_{R}^{\left( 2\right) }$ and $N_{R}^{\left( 3\right) }$. This symmetry also
distinguishes the $SU\left( 5\right) $ quintuplets $H_{i}^{\left( 2\right) }$
charged under this symmetry from the remaing $SU\left( 5\right) $
quintuplets $H_{i}^{\left( 1\right) }$, $H_{i}^{\left( 3\right) }$, neutral
under this symmetry.

\item The $Z_{2}^{\prime \prime }$ symmetry distinguishes the right handed
heavy Majorana neutrinos, odd under this symmetry from the remaining
fermionic fields, even under $Z_{2}^{\prime \prime }$. In the scalar sector,
only $\eta $ and $\zeta $ are odd under this symmetry, while the remaining
scalars are $Z_{2}^{\prime \prime }$ even.

\item The $Z_{4}$ symmetry separates the fermionic $\overline{\mathbf{5}}%
^{\left( 1\right) }$\ irrep of $SU\left( 5\right) $ belonging to the first
family from the remaining fermionic $\overline{\mathbf{5}}$ irreps of $%
SU\left( 5\right) $, neutral under this symmetry. This $Z_{4}$ symmetry also
distinguishes the $SU\left( 5\right) $ scalar singlets $\sigma $ and $\eta $
charged under $Z_{4}$ from the remaining scalar fields, neutral under this
symmetry. Furthermore, it is assumed that the the right handed heavy
Majorana neutrino are charged under $Z_{4}$ symmetry. Without the $Z_{4}$
charged $SU\left( 5\right) $ scalar singlet $\sigma $, the down quark and
electron masses would be larger by about two orders of magnitude than their
corresponding experimental values, unless one sets the corresponding Yukawa
couplings unnaturally small. It is noteworthy, that unlike in the up type
quark sector, a $\lambda ^{8}$ supression in the 11 entry of the mass
matrices for down type quarks and charged leptons is required to naturally
explain the smallness of the down quark and electron masses. The $Z_{4}$ and
the $Z_{12}$ symmetries will be crucial to achieve that $\lambda ^{8}$
supression, where $\lambda =0.225$ is one of the Wolfenstein parameters.%

\item The $Z_{12}$ symmetry shapes the hierarchical structure of the quark
mass matrices necessary to get a realistic pattern of quark masses and
mixings. Besides that, the charged lepton mass hierarchy also arises from $%
Z_{12}$ symmetry. Let us recall that due to the properties of the $Z_{N}$
groups, it follows that $Z_{12}$ is the lowest cyclic symmetry that allows
to build a ten dimensional up type quark Yukawa term with a $\frac{\chi ^{6}%
}{\Lambda ^{6}}$ insertion in a four dimensional term, crucial to get the
required $\lambda ^{6}$ supression in the 11 entry of the up type quark mass
matrix. This symmetry distinguishes the fermionic $\mathbf{10}^{\left(
3\right) }$ irrep of $SU\left( 5\right) $ corresponding to the third family,
i.e, $\Psi _{ij}^{\left( 3\right) }$, neutral under $Z_{12}$ from the
remaining fermionic fields, charged under this symmetry. It is assumed that
all fermionic $\overline{\mathbf{5}}^{\left( f\right) }$ irreps of $SU\left(
5\right) $ ($f=1,2,3$) have the same $Z_{12}$ charges, different from the $%
Z_{12}$ charge of $\overline{\mathbf{5}}^{\left( 1\right) }$. All scalars
are neutral under the $Z_{12}$ symmetry, except the $SU\left( 5\right) $
scalar singlet $\chi $. The $Z_{12}$ symmetry strongly supresses mixings of
the third generation right handed heavy Majorana neutrino $N_{R}^{\left(
3\right) }$, charged under this symmetry, with the first and second
generation ones $N_{R}^{\left( 1\right) }$ and $N_{R}^{\left( 2\right) }$,
which are $Z_{12}$ neutral. Note that the heavy Majorana neutrino $%
N_{R}^{\left( 3\right) }$ is the only fermion which is assumed to be charged
under $Z_{12}$.
\end{enumerate}

\quad We consider the following VEV pattern of the scalars fields of the
model. The VEVs of the scalars $H_{i}^{\left( f\right) }$ ($f=1,2,3$) and $%
\Sigma _{j}^{i}$ are given by: 
\begin{equation}
\left\langle H_{i}^{\left( f\right) }\right\rangle =v_{H}^{\left( f\right)
}\delta _{i5},\hspace{1.5cm}\left\langle \Sigma _{j}^{i}\right\rangle
=v_{\Sigma }\,diag\left( 1,1,1,-\frac{3}{2},-\frac{3}{2}\right) ,\hspace{%
1.5cm}f=1,2,3,
\end{equation}%
%
%
%
%
%
%
%
%
%
%
%
%
%
%
%
%
%
%
%
%
%
%
%
%
%
Furthermore, the VEV pattern for the $\Sigma $ field given above, which is
consistent with the minimization conditions of the scalar potential, follows
from the general group theory of spontaneous symmetry breakdown, as shown in
Ref. \cite{Li:1973mq}.

\quad Assuming that the hierarchy of charged fermion masses and quark mixing
matrix elements is explained by the $Z_{4}$ and $Z_{12}$ symmetries, and in
order to relate the quark masses with the quark mixing parameters, we set
the VEVs of the $SU(5)$ scalar singlets as follows:%
\begin{equation}
v_{\eta }\sim v_{\zeta }\sim v_{\chi }=v_{\sigma }=\Lambda _{GUT}=\lambda
\Lambda ,
\end{equation}%
where $\lambda =0.225$ is one of the parameters in the Wolfenstein
parametrization and $\Lambda $ corresponds to the cutoff of our model.

\quad From the properties of the $\mathbf{45}$\ dimensional irrep of $SU(5)$%
, it follows that $\Phi _{jk}^{i}$ satisfies the following relations \cite%
{Frampton:1979,Georgi:1979df}: 
\begin{equation}
\Phi _{jk}^{i}=-\Phi _{kj}^{i},\hspace{1.5cm}\sum_{i=1}^{5}\Phi _{ij}^{i}=0,%
\hspace{1.5cm}i,j,k=1,2,\cdots ,5.
\end{equation}%
%
%
%
%
%
%
%
%
%
%
%
%
%
%
%
%
%
This results in the following only allowed non-zero VEVs of $\Phi _{jk}^{i}$%
: 
\begin{equation}
\left\langle \Phi _{p5}^{p}\right\rangle =-\frac{1}{3}\left\langle \Phi
_{45}^{4}\right\rangle =v_{\Phi },\hspace{0.75cm}\left\langle \Phi
_{j5}^{i}\right\rangle =v_{\Phi }\left( \delta _{j}^{i}-4\delta
_{4}^{i}\delta _{j}^{4}\right) ,\hspace{0.75cm}i,j=1,2,3,4,5,\hspace{0.75cm}%
p=1,2,3,5.
\end{equation}%
With the above particle content, the relevant Yukawa terms invariant under
the group $\mathcal{G}$ are: 
\begin{eqnarray}
\tciLaplace _{Y} &=&\alpha _{11}\psi ^{i\left( 1\right) }H^{j\left( 3\right)
}\Psi _{ij}^{\left( 1\right) }\frac{\chi ^{6}\sigma ^{2}}{\Lambda ^{8}}%
+\beta _{11}\psi ^{i\left( 1\right) }\Phi _{i}^{jk}\Psi _{jk}^{\left(
1\right) }\frac{\chi ^{6}\sigma ^{2}}{\Lambda ^{8}}+\alpha _{22}\psi
^{i\left( 2\right) }H^{j\left( 3\right) }\Psi _{ij}^{\left( 2\right) }\frac{%
\chi ^{5}}{\Lambda ^{5}}+\beta _{22}\psi ^{i\left( 2\right) }\Phi
_{i}^{jk}\Psi _{jk}^{\left( 2\right) }\frac{\chi ^{5}}{\Lambda ^{5}}  \notag
\\
&&+\alpha _{23}\psi ^{i\left( 3\right) }H^{j\left( 3\right) }\Psi
_{ij}^{\left( 2\right) }\frac{\chi ^{5}}{\Lambda ^{5}}+\beta _{23}\psi
^{i\left( 3\right) }\Phi _{i}^{jk}\Psi _{jk}^{\left( 2\right) }\frac{\chi
^{5}}{\Lambda ^{5}}+\alpha _{32}\psi ^{i\left( 2\right) }H^{j\left( 3\right)
}\Psi _{ij}^{\left( 3\right) }\frac{\chi ^{3}}{\Lambda ^{3}}+\alpha
_{32}\psi ^{i\left( 2\right) }\Phi _{i}^{jk}\Psi _{jk}^{\left( 3\right) }%
\frac{\chi ^{3}}{\Lambda ^{3}}  \notag \\
&&+\alpha _{33}\psi ^{i\left( 3\right) }H^{j\left( 3\right) }\Psi
_{ij}^{\left( 3\right) }\frac{\chi ^{3}}{\Lambda ^{3}}+\beta _{33}\psi
^{i\left( 3\right) }\Phi _{i}^{jk}\Psi _{jk}^{\left( 3\right) }\frac{\chi
^{3}}{\Lambda ^{3}}  \notag \\
&&+\varepsilon ^{ijklp}\left\{ \gamma _{11}\Psi _{ij}^{\left( 1\right)
}H_{p}^{\left( 1\right) }\Psi _{kl}^{\left( 1\right) }\frac{\chi ^{6}}{%
\Lambda ^{6}}+\gamma _{22}\Psi _{ij}^{\left( 2\right) }H_{p}^{\left(
1\right) }\Psi _{kl}^{\left( 2\right) }\frac{\chi ^{4}}{\Lambda ^{4}}+\gamma
_{33}\Psi _{ij}^{\left( 3\right) }H_{p}^{\left( 1\right) }\Psi _{kl}^{\left(
3\right) }\right.  \notag \\
&&+\left. \gamma _{12}\Psi _{ij}^{\left( 1\right) }H_{p}^{\left( 1\right)
}\Psi _{kl}^{\left( 2\right) }\frac{\chi ^{5}}{\Lambda ^{5}}+\gamma
_{21}\Psi _{ij}^{\left( 2\right) }H_{p}^{\left( 1\right) }\Psi _{kl}^{\left(
1\right) }\frac{\chi ^{5}}{\Lambda ^{5}}+\gamma _{13}\Psi _{ij}^{\left(
1\right) }H_{p}^{\left( 1\right) }\Psi _{kl}^{\left( 3\right) }\frac{\chi
^{3}}{\Lambda ^{3}}\right.  \notag \\
&&+\left. \gamma _{31}\Psi _{ij}^{\left( 3\right) }H_{p}^{\left( 1\right)
}\Psi _{kl}^{\left( 1\right) }\frac{\chi ^{3}}{\Lambda ^{3}}+\gamma
_{23}\Psi _{ij}^{\left( 2\right) }H_{p}^{\left( 2\right) }\Psi _{kl}^{\left(
3\right) }\frac{\chi ^{2}}{\Lambda ^{2}}+\gamma _{32}\Psi _{ij}^{\left(
3\right) }H_{p}^{\left( 2\right) }\Psi _{kl}^{\left( 2\right) }\frac{\chi
^{2}}{\Lambda ^{2}}\right\}  \notag \\
&&+\left( y_{1}\overline{N}_{R}^{\left( 1\right) }N_{R}^{\left( 1\right)
c}+y_{2}\overline{N}_{R}^{\left( 2\right) }N_{R}^{\left( 2\right) c}+y_{3}%
\overline{N}_{R}^{\left( 3\right) }N_{R}^{\left( 3\right) c}\right) \frac{%
\chi ^{\ast }\chi +x_{1}\sigma ^{\ast }\sigma +x_{2}\eta ^{2}+x_{3}\zeta ^{2}%
}{\Lambda }  \notag \\
&&+\varepsilon _{11}\psi ^{i\left( 1\right) }H_{i}^{\left( 1\right)
}N_{R}^{\left( 1\right) }\frac{\chi ^{3}\eta }{\Lambda ^{4}}+\varepsilon
_{21}\psi ^{i\left( 2\right) }H_{i}^{\left( 2\right) }N_{R}^{\left( 1\right)
}\frac{\chi ^{3}\zeta }{\Lambda ^{4}}+\varepsilon _{31}\psi ^{i\left(
3\right) }H_{i}^{\left( 2\right) }N_{R}^{\left( 1\right) }\frac{\chi
^{3}\zeta }{\Lambda ^{4}}  \notag \\
&&+\varepsilon _{12}\psi ^{i\left( 1\right) }H_{i}^{\left( 2\right)
}N_{R}^{\left( 2\right) }\frac{\chi ^{3}\eta }{\Lambda ^{4}}+\varepsilon
_{22}\psi ^{i\left( 2\right) }H_{i}^{\left( 1\right) }N_{R}^{\left( 2\right)
}\frac{\chi ^{3}\zeta }{\Lambda ^{4}}+\varepsilon _{32}\psi ^{i\left(
3\right) }H_{i}^{\left( 1\right) }N_{R}^{\left( 2\right) }\frac{\chi
^{3}\zeta }{\Lambda ^{4}}  \label{ly2b}
\end{eqnarray}%
where the dimensionless couplings in Eq. (\ref{ly2b}) are $\mathcal{O}(1)$
parameters. It is worth mentioning that the terms in the first, second and
third lines of Eq. (\ref{ly2b}) contribute to the masses of the down-type
quarks and charged leptons, the terms of the fourth, fifth and sixth lines
of Eq. (\ref{ly2b}) give contributions to the up-type quark masses, while
the remaining terms generate the neutrino masses. The aforementioned Yukawa
terms do not include the interactions involving the third generation right
handed heavy Majorana neutrino $N_{R}^{\left( 3\right) }$, since they are
strongly suppressed by powers of $\frac{v_{\chi }^{6}}{\Lambda ^{6}}=\lambda
^{6}$, where $\lambda =0.225$ is one of the Wolfenstein parameters. This is
a consequence of the nontrivial $Z_{12}$ charge assignement for the third
generation right handed heavy Majorana neutrino. 
The lightest of the physical neutral scalar states of $H^{(1)}$, $H^{(2)}$, $%
H^{(3)}$ and $\Phi _{jk}^{i}$ is the SM-like 126 GeV Higgs discovered at the
LHC \cite{LHC-H-discovery}. Besides that, the resulting low energy effective
theory corresponds to a four Higgs doublet model (4HDM), with a light scalar
color octet. Note that our model is much more economical than our previous $%
SU(5)$ GUT model with $A_{4}$ flavour symmetry of Ref. \cite{Campos:2014lla}%
, whose low energy effective theory corresponds to an eight Higgs doublet
model (8HDM), with a light scalar octet. As we will show in section \ref%
{quarkmassesandmixing}, the dominant contribution to the top quark mass
mainly arises from $H^{(1)}$. 
The SM-like 126 GeV Higgs also receives its main contributions from the CP
even neutral state of the $SU(2)$ doublet part of $H^{(1)}$. 
The remaining scalars are heavy and outside the LHC reach. The large number
of free uncorrelated parameters in the scalar potential allows us to adjust
the required pattern of scalar masses. Therefore, by an appropiate choice of
the free parameters in the scalar potential, one can suppress the loop
effects of the heavy scalars contributing to certain observables. 

\section{Quark masses and mixing}

\label{quarkmassesandmixing} 
From Eq. (\ref{ly2b}) we get the following mass matrix textures for quarks: 
\begin{equation}
M_{U}=\left( 
\begin{array}{ccc}
a_{11}^{\left( U\right) }\lambda ^{6} & a_{12}^{\left( U\right) }\lambda ^{5}
& a_{13}^{\left( U\right) }\lambda ^{3} \\ 
a_{12}^{\left( U\right) }\lambda ^{5} & a_{22}^{\left( U\right) }\lambda ^{4}
& a_{23}^{\left( U\right) }\lambda ^{2} \\ 
a_{13}^{\left( U\right) }\lambda ^{3} & a_{23}^{\left( U\right) }\lambda ^{2}
& a_{33}^{\left( U\right) }%
\end{array}%
\right) \allowbreak \frac{v}{\sqrt{2}},  \label{MU}
\end{equation}%
\begin{equation}
M_{D}=\left( 
\begin{array}{ccc}
a_{11}^{\left( D\right) }\lambda ^{8} & 0 & 0 \\ 
0 & a_{22}^{\left( D\right) }\lambda ^{5} & a_{23}^{\left( D\right) }\lambda
^{5} \\ 
0 & a_{32}^{\left( D\right) }\lambda ^{3} & a_{33}^{\left( D\right) }\lambda
^{3}%
\end{array}%
\right) \frac{v}{\sqrt{2}},  \label{MD0}
\end{equation}%
Furthermore, the $\mathcal{O}(1)$ dimensionless couplings in Eqs. (\ref{MU})
and (\ref{MD0}) are given by the following relations: 
\begin{eqnarray}
a_{12}^{\left( U\right) } &=&2\sqrt{2}\left( \gamma _{12}+\gamma
_{21}\right) \frac{v_{H}^{\left( 2\right) }}{v},\hspace{1cm}a_{11}^{\left(
U\right) }=4\sqrt{2}\gamma _{11}\frac{v_{H}^{\left( 1\right) }}{v},\hspace{%
1cm}a_{13}^{\left( U\right) }=2\sqrt{2}\left( \gamma _{13}+\gamma
_{31}\right) \frac{v_{H}^{\left( 1\right) }}{v},  \notag \\
a_{23}^{\left( U\right) } &=&2\sqrt{2}\left( \gamma _{23}+\gamma
_{32}\right) \frac{v_{H}^{\left( 1\right) }}{v},\hspace{1cm}a_{22}^{\left(
U\right) }=4\sqrt{2}\gamma _{22}\frac{v_{H}^{\left( 1\right) }}{v},\hspace{%
1cm}a_{33}^{\left( U\right) }=4\sqrt{2}\gamma _{33}\frac{v_{H}^{\left(
1\right) }}{v},  \notag \\
a_{23}^{\left( D\right) } &=&\frac{1}{v}\left( \alpha _{23}v_{H}^{\left(
3\right) }+2\beta _{23}v_{\Phi }\right) ,\hspace{1cm}a_{11}^{\left( D\right)
}=\frac{1}{v}\left( \alpha _{11}v_{H}^{\left( 3\right) }+2\beta _{11}v_{\Phi
}\right) ,\hspace{1cm}a_{22}^{\left( D\right) }=\frac{1}{v}\left( \alpha
_{22}v_{H}^{\left( 3\right) }+2\beta _{22}v_{\Phi }\right) ,  \notag \\
a_{32}^{\left( D\right) } &=&\frac{1}{v}\left( \alpha _{32}v_{H}^{\left(
3\right) }+2\beta _{32}v_{\Phi }\right) ,\hspace{1cm}a_{33}^{\left( D\right)
}=\frac{1}{v}\left( \alpha _{33}v_{H}^{\left( 3\right) }+2\beta _{33}v_{\Phi
}\right) .  \label{couplings}
\end{eqnarray}%
Assuming that the hierarchy of charged fermion masses and quark mixing
matrix elements is explained by the $Z_{12}$ symmetry, we adopt an
approximate universality in the dimensionless Yukawa couplings for up type
quarks, down type quarks and charged leptons: 
\begin{eqnarray}
\gamma _{11} &=&\gamma _{1},\hspace{1cm}\gamma _{12}=\gamma _{21}=\gamma
_{1}\left( 1-\frac{\lambda ^{2}}{2}\right) ^{2},\hspace{1cm}\gamma
_{22}=\gamma _{1}\left( 1-\frac{\lambda ^{2}}{2}\right) ^{3},\hspace{1cm}%
\gamma _{13}=\gamma _{31}=-\gamma _{2}e^{-i\phi },  \notag \\
\gamma _{23} &=&\gamma _{32}=-\gamma _{2}\left( 1-\frac{\lambda ^{2}}{2}%
\right) e^{-i\phi },\hspace{1cm}\gamma _{33}=\gamma _{4}e^{-2i\phi },\hspace{%
1cm}\alpha _{ii}=\alpha _{i},\hspace{1cm}\beta _{ii}=\beta _{i},\hspace{2cm}
\notag \\
\alpha _{ij} &=&\widetilde{\alpha },\hspace{1cm}\beta _{ij}=\widetilde{\beta 
},\hspace{1cm}i\neq j,\hspace{1cm}i,j=1,2,3.  \label{universality}
\end{eqnarray}%
where $\lambda =0.225$, with $\gamma _{1}$, $\gamma _{2}$, $\widetilde{%
\alpha }$, $\widetilde{\beta }$, $\alpha _{i}$ and $\beta _{i}$ ($i=1,2,3$)\
real $\mathcal{O}(1)$ parameters. Note that exact universality in the
dimensionless quark Yukawa couplings leads to massless up, charm and strange
quarks. Consequently a breaking of universality in the quark Yukawa
couplings is required to generate these masses. 
Furthermore, for the sake of simplicity, we assume that the complex phase
responsible for CP violation in the quark sector only arises from the up
type quark sector. In addition, to simplify the analysis, we set $%
v_{H}^{\left( 1\right) }=v_{H}^{\left( 2\right) }$ and we fix $\left\vert
a_{33}^{\left( U\right) }\right\vert =1$, as suggested by the naturalness
arguments. Therefore, the up and down type quark mass matrices take the
following form: 
\begin{eqnarray}
M_{U} &=&P^{\dag }\widetilde{M}_{U}P^{\dag },\hspace{2cm}P=\left( 
\begin{array}{ccc}
1 & 0 & 0 \\ 
0 & 1 & 0 \\ 
0 & 0 & e^{i\phi }%
\end{array}%
\right) ,  \notag \\
\widetilde{M}_{U} &=&\left( 
\begin{array}{ccc}
a_{1}^{\left( U\right) }\lambda ^{6} & a_{1}^{\left( U\right) }\left( 1-%
\frac{\lambda ^{2}}{2}\right) ^{2}\lambda ^{5} & -a_{2}^{\left( U\right)
}\lambda ^{3} \\ 
a_{1}^{\left( U\right) }\left( 1-\frac{\lambda ^{2}}{2}\right) ^{2}\lambda
^{5} & a_{1}^{\left( U\right) }\left( 1-\frac{\lambda ^{2}}{2}\right)
^{3}\lambda ^{4} & -a_{2}^{\left( U\right) }\left( 1-\frac{\lambda ^{2}}{2}%
\right) \lambda ^{2} \\ 
-a_{2}^{\left( U\right) }\left( 1-\frac{\lambda ^{2}}{2}\right) \lambda ^{3}
& -a_{2}^{\left( U\right) }\left( 1-\frac{\lambda ^{2}}{2}\right) \lambda
^{2} & 1%
\end{array}%
\right) \allowbreak \frac{v}{\sqrt{2}},  \label{Mu}
\end{eqnarray}%
\begin{equation}
M_{D}=\left( 
\begin{array}{ccc}
a_{1}^{\left( D\right) }\lambda ^{8} & 0 & 0 \\ 
0 & a_{2}^{\left( D\right) }\lambda ^{5} & a_{4}^{\left( D\right) }\lambda
^{5} \\ 
0 & a_{4}^{\left( D\right) }\lambda ^{3} & a_{3}^{\left( D\right) }\lambda
^{3}%
\end{array}%
\right) \frac{v}{\sqrt{2}}.  \label{Md}
\end{equation}%
From Eqs. (\ref{Mu}) and (\ref{Md}) it follows that the up and down type
quark masses are approximatelly given by: 
\begin{eqnarray}
m_{u} &\simeq &\frac{a_{1}^{\left( U\right) }}{2}\left( 1-\frac{\lambda ^{2}%
}{2}\right) \frac{\left[ \left( a_{2}^{\left( U\right) }\right)
^{2}-a_{1}^{\left( U\right) }\left( 1-\frac{\lambda ^{2}}{2}\right) \right] 
}{1+\lambda a_{1}^{\left( U\right) }\left( 1-\frac{\lambda ^{2}}{2}\right)
^{3}}\lambda ^{8}\frac{v}{\sqrt{2}},\hspace{1cm}m_{c}\simeq \left[ 1+\lambda
a_{1}^{\left( U\right) }\left( 1-\frac{\lambda ^{2}}{2}\right) ^{3}\right] 
\frac{\lambda ^{4}v}{\sqrt{2}},\hspace{1cm}m_{t}\simeq \frac{v}{\sqrt{2}}, 
\notag \\
m_{d} &=&a_{1}^{\left( D\right) }\lambda ^{8}\frac{v}{\sqrt{2}},\hspace{1cm}%
m_{s}\simeq \frac{\left\vert a_{2}^{\left( D\right) }a_{3}^{\left( D\right)
}-\left( a_{4}^{\left( D\right) }\right) ^{2}\right\vert }{\sqrt{\left(
a_{3}^{\left( D\right) }\right) ^{2}+\left( a_{4}^{\left( D\right) }\right)
^{2}}}\lambda ^{5}\frac{v}{\sqrt{2}},\hspace{1cm}m_{b}\simeq \sqrt{\left(
a_{3}^{\left( D\right) }\right) ^{2}+\left( a_{4}^{\left( D\right) }\right)
^{2}}\lambda ^{3}\frac{v}{\sqrt{2}}.  \label{mq}
\end{eqnarray}%
The CKM quark mixing matrix is approximatelly given by: 
\begin{equation}
V_{CKM}=R_{U}^{T}PR_{D}\simeq \left( 
\begin{array}{ccc}
c & sc_{D} & ss_{D} \\ 
-sc_{U} & cc_{U}c_{D}-s_{U}s_{D}e^{i\phi } & s_{U}c_{D}e^{i\phi }+cc_{U}s_{D}
\\ 
ss_{U} & -c_{U}s_{D}e^{i\phi }-cs_{U}c_{D} & c_{U}c_{D}e^{i\phi }-cs_{U}s_{D}%
\end{array}%
\right) ,  \label{CKM}
\end{equation}%
where $c=\cos \theta $, $s=\sin \theta $, $c_{U,D}=\cos \theta _{U,D}$, $%
s_{U,D}=\sin \theta _{U,D}$ and the quark mixing angles are: 
\begin{equation}
\sin \theta \simeq -\lambda ,\hspace{2cm}\sin \theta _{U}\simeq -\frac{%
\lambda a_{2}^{\left( D\right) }\left( 1-\frac{\lambda ^{2}}{2}\right) ^{3}}{%
2a_{1}^{\left( D\right) }},\hspace{2cm}\tan 2\theta _{D}=\frac{2\left(
a_{2}^{\left( D\right) }+a_{3}^{\left( D\right) }\right) a_{4}^{\left(
D\right) }}{\left( a_{3}^{\left( D\right) }\right) ^{2}+\left( a_{4}^{\left(
D\right) }\right) ^{2}}\lambda ^{2}.  \label{quarkmixingangles}
\end{equation}%
%
%
%
%
%
%
%
%
%
%
%
%
%
%
%
%
It is noteworthy that Eqs. (\ref{mq})-(\ref{quarkmixingangles}) provide an
elegant understanding of all SM quark masses and mixing parameters in terms
of the Wolfenstein parameter $\lambda =0.225$ and of parameters of order
unity. Note that all physical parameters in the quark sector are linked with
the electroweak symmetry breaking scale $v=246$ GeV through their scalings
by powers of the Wolfenstein parameter $\lambda =0.225$, with $\mathcal{O}%
(1) $ coefficients. 
\begin{table}[tbh]
\begin{center}
\begin{tabular}{c|l|l}
\hline\hline
Observable & Model value & Experimental value \\ \hline
$m_{u}(MeV)$ & \quad $1.11$ & \quad $1.45_{-0.45}^{+0.56}$ \\ \hline
$m_{c}(MeV)$ & \quad $639$ & \quad $635\pm 86$ \\ \hline
$m_{t}(GeV)$ & \quad $172.3$ & \quad $172.1\pm 0.6\pm 0.9$ \\ \hline
$m_{d}(MeV)$ & \quad $2.9$ & \quad $2.9_{-0.4}^{+0.5}$ \\ \hline
$m_{s}(MeV)$ & \quad $57.7$ & \quad $57.7_{-15.7}^{+16.8}$ \\ \hline
$m_{b}(GeV)$ & \quad $2.82$ & \quad $2.82_{-0.04}^{+0.09}$ \\ \hline
$\bigl|V_{ud}\bigr|$ & \quad $0.974$ & \quad $0.97427\pm 0.00015$ \\ \hline
$\bigl|V_{us}\bigr|$ & \quad $0.22516$ & \quad $0.22534\pm 0.00065$ \\ \hline
$\bigl|V_{ub}\bigr|$ & \quad $0.00353$ & \quad $%
0.00351_{-0.00014}^{+0.00015} $ \\ \hline
$\bigl|V_{cd}\bigr|$ & \quad $0.22502$ & \quad $0.22520\pm 0.00065$ \\ \hline
$\bigl|V_{cs}\bigr|$ & \quad $0.97348$ & \quad $0.97344\pm 0.00016$ \\ \hline
$\bigl|V_{cb}\bigr|$ & \quad $0.0412$ & \quad $0.0412_{-0.0005}^{+0.0011}$
\\ \hline
$\bigl|V_{td}\bigr|$ & \quad $0.00860$ & \quad $%
0.00867_{-0.00031}^{+0.00029} $ \\ \hline
$\bigl|V_{ts}\bigr|$ & \quad $0.0404$ & \quad $0.0404_{-0.0005}^{+0.0011}$
\\ \hline
$\bigl|V_{tb}\bigr|$ & \quad $0.999145$ & \quad $%
0.999146_{-0.000046}^{+0.000021}$ \\ \hline
$J$ & \quad $2.96\times 10^{-5}$ & \quad $(2.96_{-0.16}^{+0.20})\times
10^{-5}$ \\ \hline
$\delta $ & \quad $68^{\circ }$ & \quad $68^{\circ }$ \\ \hline\hline
\end{tabular}%
\end{center}
\caption{Model and experimental values of the quark masses and CKM
parameters.}
\label{Observables}
\end{table}

\quad To describe the quark masses and mixing, we have 9 parameters, i.e, $%
\lambda $, $a_{1}^{\left( U\right) }$, $a_{2}^{\left( U\right) }$, $%
a_{33}^{\left( U\right) }$, $a_{3}^{\left( D\right) }$, $a_{4}^{\left(
D\right) }$, $e_{D}$, $f_{D}$ and the phases $\phi $, while the
corresponding number of observables in the quark sector is 10. Note that the
parameters $\lambda $ and $a_{33}^{\left( U\right) }$ are fixed while the
remaining 7 parameters are fitted to reproduce the 6 quark masses and 4
quark mixing parameters. The results shown in Table \ref{Observables}
correspond to the following best-fit values: 
\begin{eqnarray}
a_{1}^{\left( U\right) } &\simeq &2.05,\hspace{1cm}a_{2}^{\left( U\right)
}\simeq 0.75,\hspace{1cm}a_{1}^{\left( D\right) }\simeq 2.51,\hspace{1cm}%
a_{2}^{\left( D\right) }\simeq -0.31,\hspace{1cm}  \notag \\
a_{3}^{\left( D\right) } &\simeq &1.26,\hspace{1cm}a_{4}^{\left( D\right)
}\simeq -0.65,\hspace{1cm}\phi \simeq -90.24^{\circ },
\end{eqnarray}

The obtained and experimental values of the observables in the quark sector
are shown in Table \ref{Observables}. The experimental values of the quark
masses, which are given at the $M_{Z}$ scale, have been taken from Ref. \cite%
{Bora:2012tx} (which are similar to those in \cite{Xing:2007fb}), whereas
the experimental values of the CKM matrix elements, the Jarlskog invariant $%
J $ and the CP violating phase $\delta $ are taken from Ref. \cite{PDG}. As
seen from Table \ref{Observables}, the quark masses and CKM parameters are
in excellent agreement with the experimental data. The agreement of our
model with the experimental data is as good as in the models of Refs. \cite%
{Branco2010,CarcamoHernandez:2010im,Hernandez:2013hea,Hernandez:2014vta,Hernandez:2014a}%
, and better than many others approaches \cite%
{Chen:2007,Fritzsch,Xing,FX,Matsuda,Zhou,Carcamo,Xing2010,Branco2012,CarcamoHernandez:2012xy,Bhattacharyya:2012pi,King:2013hj,Hernandez:2014a,Vien:2014ica}%
. 

\newpage

\section{Lepton masses and mixing}

\label{leptonmassmix} Using Eq. (\ref{ly2b}) it follows that the charged
lepton mass matrix is given by:

\begin{equation}
M_{l}=\left( 
\begin{array}{ccc}
a_{11}^{\left( l\right) }\lambda ^{8} & 0 & 0 \\ 
0 & a_{22}^{\left( l\right) }\lambda ^{5} & a_{23}^{\left( l\right) }\lambda
^{3} \\ 
0 & a_{32}^{\left( l\right) }\lambda ^{5} & a_{33}^{\left( l\right) }\lambda
^{3}%
\end{array}%
\right) \frac{v}{\sqrt{2}},  \label{Ml}
\end{equation}

where:

\begin{eqnarray}
a_{32}^{\left( l\right) } &=&\frac{1}{v}\left( \alpha _{23}v_{S}-6\beta
_{23}v_{\Phi }\right) ,\hspace{2cm}a_{11}^{\left( l\right) }=\frac{1}{v}%
\left( \alpha _{11}v_{S}-6\beta _{11}v_{\Phi }\right) ,  \notag \\
a_{23}^{\left( l\right) } &=&\frac{1}{v}\left( \alpha _{32}v_{S}-6\beta
_{32}v_{\Phi }\right) ,\hspace{2cm}a_{22}^{\left( l\right) }=\frac{1}{v}%
\left( \alpha _{22}v_{S}-6\beta _{22}v_{\Phi }\right) ,  \notag \\
a_{33}^{\left( l\right) } &=&\frac{1}{v}\left( \alpha _{33}v_{S}-6\beta
_{33}v_{\Phi }\right) ,  \label{couplingsleptons}
\end{eqnarray}

where the dimensionless couplings in Eq. (\ref{Ml}) are $\mathcal{O}(1)$
parameters. From Eq. (\ref{universality}), it follows that the mass matrix
for charged leptons can be rewritten as follows:

\begin{equation}
M_{l}=\left( 
\begin{array}{ccc}
a_{1}^{\left( l\right) }\lambda ^{8} & 0 & 0 \\ 
0 & a_{2}^{\left( l\right) }\lambda ^{5} & a_{4}^{\left( l\right) }\lambda
^{3} \\ 
0 & a_{4}^{\left( l\right) }\lambda ^{5} & a_{3}^{\left( l\right) }\lambda
^{3}%
\end{array}%
\right) \frac{v}{\sqrt{2}}.  \label{Mls}
\end{equation}

The matrix $M_{l}M_{l}^{T}$ is diagonalized by a rotation matrix $R_{l}$
according to: 
\begin{eqnarray}
R_{l}^{T}M_{l}M_{l}^{T}R_{l} &=&diag\left( m_{e},m_{\mu },m_{\tau }\right) ,%
\hspace{1cm}R_{l}=\left( 
\begin{array}{ccc}
1 & 0 & 0 \\ 
0 & \cos \theta _{l} & -\sin \theta _{l} \\ 
0 & \sin \theta _{l} & \cos \theta _{l}%
\end{array}%
\right) ,  \notag \\
\tan \theta _{l} &\simeq &-\frac{a_{4}^{\left( l\right) }}{a_{3}^{\left(
l\right) }},\hspace{1cm}\cos \theta _{l}\simeq \frac{a_{3}^{\left( l\right) }%
}{\sqrt{\left( a_{3}^{\left( l\right) }\right) ^{2}+\left( a_{4}^{\left(
l\right) }\right) ^{2}}},\hspace{1cm}\sin \theta _{l}\simeq -\frac{%
a_{4}^{\left( l\right) }}{\sqrt{\left( a_{3}^{\left( l\right) }\right)
^{2}+\left( a_{4}^{\left( l\right) }\right) ^{2}}},  \label{Rl}
\end{eqnarray}

where, from Eq. (\ref{Mls}) it follows that the charged lepton masses are
approximatelly given by:

\begin{equation}
m_{e}=a_{1}^{\left( l\right) }\lambda ^{8}\frac{v}{\sqrt{2}},\hspace{1cm}%
m_{\mu }\simeq \frac{\left\vert a_{2}^{\left( l\right) }a_{3}^{\left(
l\right) }-\left( a_{4}^{\left( l\right) }\right) ^{2}\right\vert }{\sqrt{%
\left( a_{3}^{\left( l\right) }\right) ^{2}+\left( a_{4}^{\left( l\right)
}\right) ^{2}}}\lambda ^{5}\frac{v}{\sqrt{2}},\hspace{1cm}m_{\tau }\simeq 
\sqrt{\left( a_{3}^{\left( l\right) }\right) ^{2}+\left( a_{4}^{\left(
l\right) }\right) ^{2}}\lambda ^{3}\frac{v}{\sqrt{2}}.  \label{ml}
\end{equation}%
Note the remarkable feature that charged lepton masses are connected with
the electroweak symmetry breaking scale $v=246$ GeV through their power
dependence on the Wolfenstein parameter $\lambda =0.225$, with $\mathcal{O}%
(1)$ coefficients. This remarkable feature, which is also presented in the
quark sector, is due to the fact that down type quarks and charged leptons
are members of the same fermionic multiplets of the GUT group. Furthermore,
it is noteworthy that unlike in the down type quark sector, the mixing angle 
$\theta _{l}$ in the charged lepton sector is large, which gives rise to an
important contribution to the leptonic mixing matrix, coming from the mixing
of charged leptons. 

\quad Since the four $SU\left( 5\right) $ scalar singlets acquiring VEVs at
the GUT scale have Yukawa interactions with the right handed Majorana
neutrinos, the right handed Majorana neutrino masses have GUT scale values.
Consequently, the entries of the diagonal heavy Majorana neutrino mass
matrix satisfy $\left( M_{R}\right) _{ii}>>v$, implying that the light
neutrino masses are generated via type I seesaw mechanism. Then, from Eq. (%
\ref{ly2b}) it follows that the neutrino mass matrix is given by:

\begin{equation}
M_{\nu }=\left( 
\begin{array}{cc}
O_{3\times 3} & M_{\nu }^{D} \\ 
\left( M_{\nu }^{D}\right) ^{T} & M_{R}%
\end{array}%
\right) ,\hspace{2cm}M_{\nu }^{D}=\allowbreak \left( 
\begin{array}{ccc}
A & D & 0 \\ 
B & E & 0 \\ 
C & F & 0%
\end{array}%
\right) ,\hspace{2cm}M_{R}=\left( 
\begin{array}{ccc}
M_{1} & 0 & 0 \\ 
0 & M_{2} & 0 \\ 
0 & 0 & M_{3}%
\end{array}%
\right) ,  \label{neutrinomatrix}
\end{equation}
where: 
\begin{eqnarray}
A &=&\varepsilon _{11}\lambda ^{3}v_{H}^{\left( 1\right) }\frac{v_{\eta }}{%
\Lambda },\hspace{2cm}D=\allowbreak \varepsilon _{12}\lambda
^{3}v_{H}^{\left( 2\right) }\frac{v_{\eta }}{\Lambda },\hspace{2cm}%
B=\allowbreak \varepsilon _{21}\lambda ^{3}v_{H}^{\left( 2\right) }\frac{%
v_{\zeta }}{\Lambda },  \notag \\
E &=&\allowbreak \varepsilon _{22}\lambda ^{3}v_{H}^{\left( 1\right) }\frac{%
v_{\zeta }}{\Lambda },\hspace{2cm}C=\allowbreak \varepsilon _{31}\lambda
^{3}v_{H}^{\left( 2\right) }\frac{v_{\zeta }}{\Lambda },\hspace{2cm}%
F=\allowbreak \varepsilon _{32}\lambda ^{3}v_{H}^{\left( 1\right) }\frac{%
v_{\eta }}{\Lambda },  \notag \\
M_{i} &=&y_{i}\frac{v_{\chi }^{2}+x_{1}v_{\sigma }^{2}+x_{2}v_{\eta
}^{2}+x_{3}v_{\zeta }^{2}}{\Lambda },\hspace{2cm}i=1,2,3.
\label{neutrinocouplings}
\end{eqnarray}

Therefore the light neutrino mass matrix takes the following form:

\begin{equation}
M_{L}=M_{\nu }^{D}M_{R}^{-1}\left( M_{\nu }^{D}\right) ^{T}=\left( 
\begin{array}{ccc}
W^{2} & WX\cos \varphi & WY\cos \left( \varphi -\varrho \right) \\ 
WX\cos \varphi & X^{2} & XY\cos \varrho \\ 
WY\cos \left( \varphi -\varrho \right) & XY\cos \varrho & Y^{2}%
\end{array}%
\right) ,
\end{equation}
where: 
\begin{eqnarray}
W &=&\left\vert \overrightarrow{W}\right\vert =\allowbreak \sqrt{\frac{A^{2}%
}{M_{1}}+\frac{D^{2}}{M_{2}}},\hspace{2cm}X=\left\vert \overrightarrow{X}%
\right\vert =\sqrt{\frac{B^{2}}{M_{1}}+\frac{E^{2}}{M_{2}}},\hspace{2cm}%
Y=\left\vert \overrightarrow{Y}\right\vert =\sqrt{\frac{C^{2}}{M_{1}}+\frac{%
F^{2}}{M_{2}}},  \notag \\
\overrightarrow{W} &=&\left( \frac{A}{\sqrt{M_{1}}},\frac{D}{\sqrt{M_{2}}}%
\right) ,\hspace{2cm}\overrightarrow{X}=\left( \frac{B}{\sqrt{M_{1}}},\frac{E%
}{\sqrt{M_{2}}}\right) ,\hspace{2cm}\overrightarrow{Y}=\left( \frac{C}{\sqrt{%
M_{1}}},\frac{F}{\sqrt{M_{2}}}\right) ,  \notag \\
\cos \varphi &=&\frac{\overrightarrow{W}\cdot \overrightarrow{X}}{\left\vert 
\overrightarrow{W}\right\vert \left\vert \overrightarrow{X}\right\vert },%
\hspace{2cm}\cos \left( \varphi -\varrho \right) =\frac{\overrightarrow{W}%
\cdot \overrightarrow{Y}}{\left\vert \overrightarrow{W}\right\vert
\left\vert \overrightarrow{Y}\right\vert },\hspace{2cm}\cos \varrho =\frac{%
\overrightarrow{X}\cdot \overrightarrow{Y}}{\left\vert \overrightarrow{X}%
\right\vert \left\vert \overrightarrow{Y}\right\vert }.
\end{eqnarray}

To simplify the analysis, we set $\varphi =\varrho $, obtaining that the
light neutrino mass matrix is given by:

\begin{equation}
M_{L}=\left( 
\begin{array}{ccc}
W^{2} & \kappa WX & WY \\ 
\kappa WX & X^{2} & \kappa XY \\ 
WY & \kappa XY & Y^{2}%
\end{array}%
\right) ,\hspace{2cm}\kappa =\cos \varphi .
\end{equation}

Assuming that the neutrino Yukawa couplings are real, we find that for the
normal (NH) and inverted (IH) mass hierarchies, the light neutrino mass
matrix is diagonalized by a rotation matrix $R_{\nu }$, according to:

\begin{eqnarray}
R_{\nu }^{T}M_{L}R_{\nu } &=&\left( 
\begin{array}{ccc}
0 & 0 & 0 \\ 
0 & m_{\nu _{2}} & 0 \\ 
0 & 0 & m_{\nu _{3}}%
\end{array}%
\right) \allowbreak ,\hspace{1cm}R_{\nu }=\left( 
\begin{array}{ccc}
-\frac{Y}{\sqrt{W^{2}+Y^{2}}} & \frac{W}{\sqrt{W^{2}+Y^{2}}}\sin \theta
_{\nu } & \frac{W}{\sqrt{W^{2}+Y^{2}}}\cos \theta _{\nu } \\ 
0 & \cos \theta _{\nu } & -\sin \theta _{\nu } \\ 
\frac{W}{\sqrt{W^{2}+Y^{2}}} & \frac{Y}{\sqrt{W^{2}+Y^{2}}}\sin \theta _{\nu
} & \frac{Y}{\sqrt{W^{2}+Y^{2}}}\cos \theta _{\nu }%
\end{array}%
\right) ,\hspace{1cm}\mbox{for NH}  \label{NeutrinomassNH} \\
\tan \theta _{\nu } &=&-\sqrt{\frac{m_{3}-X^{2}}{X^{2}-m_{2}}}\allowbreak ,%
\hspace{1cm}m_{\nu _{1}}=0,\hspace{1cm}m_{\nu _{2,3}}=\frac{W^{2}+X^{2}+Y^{2}%
}{2}\mp \frac{\sqrt{\left( W^{2}-X^{2}+Y^{2}\right) ^{2}-4\kappa
^{2}X^{2}\left( W^{2}+Y^{2}\right) }}{2}.  \notag
\end{eqnarray}%
\begin{eqnarray}
R_{\nu }^{T}M_{L}R_{\nu } &=&\left( 
\begin{array}{ccc}
m_{\nu _{1}} & 0 & 0 \\ 
0 & m_{\nu _{2}} & 0 \\ 
0 & 0 & 0%
\end{array}%
\right) \allowbreak ,\hspace{1cm}R_{\nu }=\left( 
\begin{array}{ccc}
\frac{W}{\sqrt{W^{2}+Y^{2}}} & -\frac{Y}{\sqrt{W^{2}+Y^{2}}}\sin \theta
_{\nu } & -\frac{Y}{\sqrt{W^{2}+Y^{2}}}\cos \theta _{\nu } \\ 
0 & \cos \theta _{\nu } & -\sin \theta _{\nu } \\ 
\frac{Y}{\sqrt{W^{2}+Y^{2}}} & \frac{W}{\sqrt{W^{2}+Y^{2}}}\sin \theta _{\nu
} & \frac{W}{\sqrt{W^{2}+Y^{2}}}\cos \theta _{\nu }%
\end{array}%
\right) \allowbreak ,\hspace{1cm}\mbox{for IH}  \label{NeutrinomassIH} \\
\allowbreak \tan \theta _{\nu } &=&-\sqrt{\frac{m_{2}-X^{2}}{X^{2}-m_{1}}}%
\allowbreak ,\hspace{1cm}m_{\nu _{1,2}}=\frac{W^{2}+X^{2}+Y^{2}}{2}\mp \frac{%
1}{2}\sqrt{\left( W^{2}-X^{2}+Y^{2}\right) ^{2}-4\kappa ^{2}X^{2}\left(
W^{2}+Y^{2}\right) },\hspace{1cm}m_{\nu _{3}}=0.  \notag
\end{eqnarray}
It is noteworthy that the smallness of the active neutrinos masses is a
consequence of their scaling with the inverse of the large Majorana neutrino
masses, as expected from the type I seesaw mechanism implemented in our
model.

\quad With the rotation matrices in the charged lepton sector $R_{l}$, given
by Eq. (\ref{Rl}), and in the neutrino sector $R_{\nu }$, given by Eqs. (\ref%
{NeutrinomassNH}) and (\ref{NeutrinomassIH}) for NH and IH, respectively, we
find the PMNS mixing matrix:

\begin{equation}
U=R_{l}^{T}R_{\nu }=\left\{ 
\begin{array}{l}
\left( 
\begin{array}{ccc}
-\frac{Y}{\sqrt{W^{2}+Y^{2}}} & \frac{W}{\sqrt{W^{2}+Y^{2}}}\sin \theta
_{\nu } & \frac{W}{\sqrt{W^{2}+Y^{2}}}\cos \theta _{\nu } \\ 
&  &  \\ 
\frac{W}{\sqrt{W^{2}+Y^{2}}}\sin \theta _{l} & \cos \theta _{l}\cos \theta
_{\nu }+\frac{Y}{\sqrt{W^{2}+Y^{2}}}\sin \theta _{l}\sin \theta _{\nu } & 
\frac{Y}{\sqrt{W^{2}+Y^{2}}}\cos \theta _{\nu }\sin \theta _{l}-\cos \theta
_{l}\sin \theta _{\nu } \\ 
&  &  \\ 
\frac{W}{\sqrt{W^{2}+Y^{2}}}\cos \theta _{l} & \frac{Y}{\sqrt{W^{2}+Y^{2}}}%
\cos \theta _{l}\sin \theta _{\nu }-\cos \theta _{\nu }\sin \theta _{l} & 
\sin \theta _{l}\sin \theta _{\nu }+\frac{Y}{\sqrt{W^{2}+Y^{2}}}\cos \theta
_{l}\cos \theta _{\nu }%
\end{array}%
\right) \allowbreak \ \ \ \ \ \mbox{for NH}, \\ 
\\ 
\left( 
\begin{array}{ccc}
\frac{W}{\sqrt{W^{2}+Y^{2}}} & -\frac{Y}{\sqrt{W^{2}+Y^{2}}}\sin \theta
_{\nu } & -\frac{Y}{\sqrt{W^{2}+Y^{2}}}\cos \theta _{\nu } \\ 
&  &  \\ 
\frac{Y}{\sqrt{W^{2}+Y^{2}}}\sin \theta _{l} & \cos \theta _{l}\cos \theta
_{\nu }+\frac{W}{\sqrt{W^{2}+Y^{2}}}\sin \theta _{\nu }\sin \theta _{l} & 
\frac{W}{\sqrt{X^{2}+Y^{2}}}\sin \theta _{l}\cos \theta _{\nu }-\cos \theta
_{l}\sin \theta _{\nu } \\ 
&  &  \\ 
\frac{Y}{\sqrt{W^{2}+Y^{2}}}\cos \theta _{l} & \frac{W}{\sqrt{W^{2}+Y^{2}}}%
\sin \theta _{\nu }\cos \theta _{l}-\cos \theta _{\nu }\sin \theta _{l} & 
\sin \theta _{l}\sin \theta _{\nu }+\frac{W}{\sqrt{W^{2}+Y^{2}}}\cos \theta
_{l}\cos \theta _{\nu }%
\end{array}%
\right) \allowbreak \ \ \ \ \ \mbox{for IH}.%
\end{array}%
\right.
\end{equation}

From the standard parametrization of the leptonic mixing matrix, it follows
that the lepton mixing angles for NH and IH, respectively, are:

\begin{eqnarray}
\sin ^{2}\theta _{12} &=&\frac{W^{2}\sin ^{2}\theta _{\nu }}{Y^{2}+\left(
1-\cos ^{2}\theta _{\nu }\right) W^{2}},\hspace{1cm}\hspace{1cm}\sin
^{2}\theta _{13}=\frac{W^{2}\cos ^{2}\theta _{\nu }}{W^{2}+Y^{2}},  \notag \\
\sin ^{2}\theta _{23} &=&\frac{\left( \sqrt{W^{2}+Y^{2}}\sin \theta _{\nu
}\cos \theta _{l}-Y\cos \theta _{\nu }\sin \theta _{l}\right) ^{2}}{\left(
1-\cos ^{2}\theta _{\nu }\right) W^{2}+Y^{2}}\ ,\ \ \ \ \ \ \ \ \mbox{for NH}
\label{mixinganglesNH}
\end{eqnarray}

\begin{eqnarray}
\sin ^{2}\theta _{12} &=&\frac{Y^{2}\sin ^{2}\theta _{\nu }}{W^{2}+\left(
1-\cos ^{2}\theta _{\nu }\right) Y^{2}},\hspace{1cm}\hspace{1cm}\sin
^{2}\theta _{13}=\frac{Y^{2}\cos ^{2}\theta _{\nu }}{W^{2}+Y^{2}},  \notag \\
\sin ^{2}\theta _{23} &=&\frac{\left( \sqrt{W^{2}+Y^{2}}\sin \theta _{\nu
}\cos \theta _{l}-W\cos \theta _{\nu }\sin \theta _{l}\right) ^{2}}{\left(
1-\cos ^{2}\theta _{\nu }\right) Y^{2}+W^{2}}\ ,\ \ \ \ \ \ \ \ \mbox{for IH}
\label{mixingnaglesIH}
\end{eqnarray}

Varying the lepton sector model parameters $a_{i}^{\left( l\right) }$ ($%
i=1,2,3,4$), $\kappa $, $W$, $X$ and $Y$, we fitted the charged lepton
masses, the neutrino mass squared splitings $\Delta m_{21}^{2}$, $\Delta
m_{31}^{2}$\ (note that we define $\Delta m_{ij}^{2}=m_{i}^{2}-m_{j}^{2}$)
and the leptonic mixing angles $\sin ^{2}\theta _{12}$, $\sin ^{2}\theta
_{13}$ and $\sin ^{2}\theta _{23}$\ to their experimental values for NH and
IH. The results shown in Table \ref{Observables0} correspond to the
following best-fit values: 
\begin{eqnarray}
\kappa &\simeq &0.45,\hspace{1cm}W\simeq 0.13eV^{\frac{1}{2}},\hspace{1cm}%
X\simeq 0.11eV^{\frac{1}{2}},\hspace{1cm}Y\simeq 0.18eV^{\frac{1}{2}}, 
\notag \\
a_{1}^{\left( l\right) } &\simeq &0.42,\hspace{1cm}a_{2}^{\left( l\right)
}\simeq 1.39,\hspace{1cm}a_{3}^{\left( l\right) }\simeq 0.77,\hspace{1cm}%
a_{4}^{\left( l\right) }\simeq 0.42,\ \ \ \ \ \ \ \ \mbox{for NH}
\label{ParameterfitNH}
\end{eqnarray}%
\begin{eqnarray}
\kappa &\simeq &4.03\times 10^{-3},\hspace{1cm}W\simeq 0.18eV^{\frac{1}{2}},%
\hspace{1cm}X\simeq 0.22eV^{\frac{1}{2}},\hspace{1cm}Y\simeq 0.13eV^{\frac{1%
}{2}},  \notag \\
a_{1}^{\left( l\right) } &\simeq &0.42,\hspace{1cm}a_{2}^{\left( l\right)
}\simeq 1.38,\hspace{1cm}a_{3}^{\left( l\right) }\simeq 0.78,\hspace{1cm}%
a_{4}^{\left( l\right) }\simeq 0.42,\ \ \ \ \ \ \ \ \mbox{for IH}
\label{ParameterfitIH}
\end{eqnarray}%
\begin{table}[tbh]
\begin{center}
\begin{tabular}{c|l|l}
\hline\hline
Observable & Model value & Experimental value \\ \hline
$m_{e}(MeV)$ & \quad $0.487$ & \quad $0.487$ \\ \hline
$m_{\mu }(MeV)$ & \quad $102.8$ & \quad $102.8\pm 0.0003$ \\ \hline
$m_{\tau }(GeV)$ & \quad $1.75$ & \quad $1.75\pm 0.0003$ \\ \hline
$\Delta m_{21}^{2}$($10^{-5}$eV$^{2}$) (NH) & \quad $7.60$ & \quad $%
7.60_{-0.18}^{+0.19}$ \\ \hline
$\Delta m_{31}^{2}$($10^{-3}$eV$^{2}$) (NH) & \quad $2.48$ & \quad $%
2.48_{-0.07}^{+0.05}$ \\ \hline
$\sin ^{2}\theta _{12}$ (NH) & \quad $0.323$ & \quad $0.323\pm 0.016$ \\ 
\hline
$\sin ^{2}\theta _{23}$ (NH) & \quad $0.567$ & \quad $%
0.567_{-0.128}^{+0.032} $ \\ \hline
$\sin ^{2}\theta _{13}$ (NH) & \quad $0.0234$ & \quad $0.0234\pm 0.0020$ \\ 
\hline
$\Delta m_{21}^{2}$($10^{-5}$eV$^{2}$) (IH) & \quad $7.60$ & \quad $%
7.60_{-0.18}^{+0.19}$ \\ \hline
$\Delta m_{13}^{2}$($10^{-3}$eV$^{2}$) (IH) & \quad $2.38$ & \quad $%
2.48_{-0.06}^{+0.05}$ \\ \hline
$\sin ^{2}\theta _{12}$ (IH) & \quad $0.323$ & \quad $0.323\pm 0.016$ \\ 
\hline
$\sin ^{2}\theta _{23}$ (IH) & \quad $0.0573$ & \quad $%
0.573_{-0.043}^{+0.025}$ \\ \hline
$\sin ^{2}\theta _{13}$ (IH) & \quad $0.0240$ & \quad $0.0240\pm 0.0019$ \\ 
\hline
\end{tabular}%
\end{center}
\par
\caption{Model and experimental values of the charged lepton masses,
neutrino mass squared splittings and leptonic mixing parameters for the
normal (NH) and inverted (IH) mass hierarchies.}
\label{Observables0}
\end{table}
Using the best-fit values given above, we get for NH and IH, respectively,
the following neutrino masses: 
\begin{equation}
m_{1}=0,\hspace{1cm}m_{2}\approx 9\mbox{meV},\hspace{1cm}m_{3}\approx 50%
\mbox{meV},\ \ \ \ \ \ \ \ \mbox{for NH}  \label{neutrinomassesNH}
\end{equation}%
\begin{equation}
m_{1}\approx 49\mbox{meV},\hspace{1cm}m_{2}\approx 50\mbox{meV},\hspace{1cm}%
m_{3}=0,\ \ \ \ \ \ \ \ \mbox{for IH}  \label{neutrinomassesIH}
\end{equation}%
The obtained and experimental values of the observables in the lepton sector
are shown in Table \ref{Observables0}. The experimental values of the
charged lepton masses, which are given at the $M_{Z}$ scale, have been taken
from Ref. \cite{Bora:2012tx} (which are similar to those in \cite%
{Xing:2007fb}), whereas the experimental values of the neutrino mass squared
splittings and leptonic mixing angles for both normal (NH) and inverted (IH)
mass hierarchies, are taken from Ref. \cite{Forero:2014bxa}. The obtained
charged lepton masses, neutrino mass squared splittings and lepton mixing
angles are in excellent agreement with the experimental data for both normal
and inverted neutrino mass hierarchies. Let us remind that for the sake of
simplicity, we assumed all leptonic parameters to be real, but a
non-vanishing CP violating phase in the PMNS mixing matrix can be generated
by making one of the entries of the neutrino mass matrix of Eq. (\ref%
{neutrinomatrix}) to be complex.

\section{Conclusions}

\label{conclusions} We proposed a model based on the $SU(5)$ grand
unification with an extra $Z_{2}\otimes Z_{2}^{\prime }\otimes Z_{2}^{\prime
\prime }\otimes Z_{4}\otimes Z_{12}$ flavor symmetry, that successfully
accounts for the SM fermion masses and mixings. The model has in total 17
effective free parameters, from which 2 are fixed and 15 are fitted to
reproduce the experimental values of 18 observables, i.e., 9 charged fermion
masses, 2 neutrino mass squared splittings, 3 lepton mixing parameters, 3
quark mixing angles and 1 CP violating phase of the CKM quark mixing matrix.
One of the two fixed parameters is identified with the Wolfenstein one and
the other one is set to one as suggested by the naturalness arguments. The
observed quark mass and mixing hierarchy is caused by the $Z_{4}$ and $%
Z_{12} $ symmetries, which are broken at very high scale by the $SU(5)$
scalar singlets $\sigma $ and $\chi $, respectively charged under these
symmetries, and which acquire VEVs at the GUT scale. The active neutrino
masses of the model are generated via type I seesaw mechanism with three
heavy Majorana neutrinos. The smallness of the active neutrino masses is
attributed to their scaling with inverse powers of the large Majorana
neutrino masses. The model predictions for the observables in both quark and
lepton sectors are in excellent agreement with the experimental data. \label%
{Summary}

\section*{Acknowledgments}

This work was partially supported by Fondecyt (Chile), Grants No. 11130115,
No. 1150792 and No. 1140390 and by DGIP internal Grant No. 111458.


\newpage

\begin{thebibliography}{99}
\bibitem{LHC-H-discovery} G. Aad et al. [ATLAS Collaboration], Phys. Lett. B 
\textbf{716}, 1 (2012); S. Chatrchyan et al. [CMS Collaboration], Phys.
Lett. B \textbf{716}, 30 (2012).

\bibitem{SM} S.L. Glashow, Nucl. Phys. 22, 579 (1961); S. Weinberg, Phys.
Rev. Lett. \textbf{19}, 1264 (1967); A. Salam, in \textit{Elementary
Particle Theory: Relativistic Groups and Analyticity (Nobel Symposium No. 8)}%
, edited by N.Svartholm (Almqvist and Wiksell, Stockholm, 1968), p. 367.

\bibitem{PDG} J. Beringer et al. (Particle Data Group), Phys.\ Rev.\ D 
\textbf{86} 010001 (2012).

\bibitem{An:2012eh} F.~P.~An \textit{et al.} [DAYA-BAY Collaboration], 
Phys.\ Rev.\ Lett.\ \textbf{108}, 171803 (2012) [arXiv:1203.1669 [hep-ex]].

\bibitem{Abe:2011sj} K.~Abe \textit{et al.} [T2K Collaboration], 
Phys.\ Rev.\ Lett.\ \textbf{107}, 041801 (2011) [arXiv:1106.2822 [hep-ex]].

\bibitem{Adamson:2011qu} P.~Adamson \textit{et al.} [MINOS Collaboration], 
Phys.\ Rev.\ Lett.\ \textbf{107}, 181802 (2011) [arXiv:1108.0015 [hep-ex]].

\bibitem{Abe:2011fz} Y.~Abe \textit{et al.} [DOUBLE-CHOOZ Collaboration], 
Phys.\ Rev.\ Lett.\ \textbf{108}, 131801 (2012) [arXiv:1112.6353 [hep-ex]].

\bibitem{Ahn:2012nd} J.~K.~Ahn \textit{et al.} [RENO Collaboration], 
Phys.\ Rev.\ Lett.\ \textbf{108}, 191802 (2012) [arXiv:1204.0626 [hep-ex]].


\bibitem{Forero:2014bxa} D.~V.~Forero, M.~Tortola and J.~W.~F.~Valle, 
arXiv:1405.7540 [hep-ph]. 

\bibitem{Fritzsch:1999ee} H.~Fritzsch and Z.~-z.~Xing, 
Prog.\ Part.\ Nucl.\ Phys.\ \textbf{45}, 1 (2000) [hep-ph/9912358].

\bibitem{Altarelli:2002hx} G.~Altarelli and F.~Feruglio, 
Springer Tracts Mod.\ Phys.\ \textbf{190}, 169 (2003) [hep-ph/0206077].

\bibitem{Altarelli:2010gt} G.~Altarelli and F.~Feruglio, 
Rev.\ Mod.\ Phys.\ \textbf{82} 2701 (2010) [arXiv:1002.0211 [hep-ph]]. 


\bibitem{Ishimori:2010au} H.~Ishimori, T.~Kobayashi, H.~Ohki, Y.~Shimizu,
H.~Okada and M.~Tanimoto, 
Prog.\ Theor.\ Phys.\ Suppl.\ \textbf{183}, 1 (2010) [arXiv:1003.3552
[hep-th]]. 

\bibitem{FlavorSymmRev} 
S.~F.~King, A.~Merle, S.~Morisi, Y.~Shimizu and M.~Tanimoto, 
New J.\ Phys.\ \textbf{16}, 045018 (2014) [arXiv:1402.4271 [hep-ph]].

\bibitem{Textures} 
R.~Barbieri, G.~R.~Dvali, A.~Strumia, Z.~Berezhiani and L.~J.~Hall, 
Nucl.\ Phys.\ B \textbf{432}, 49 (1994) [hep-ph/9405428]; R.~G.~Roberts,
A.~Romanino, G.~G.~Ross and L.~Velasco-Sevilla, 
Nucl.\ Phys.\ B \textbf{615} (2001) 358 [hep-ph/0104088]; E.~Ma and
G.~Rajasekaran, 
Phys.\ Rev.\ D \textbf{64}, 113012 (2001) [hep-ph/0106291]; 
K.~S.~Babu, E.~Ma and J.~W.~F.~Valle, 
Phys.\ Lett.\ B \textbf{552}, 207 (2003) [hep-ph/0206292]; 
G.~C.~Branco, M.~N.~Rebelo and J.~I.~Silva-Marcos, 
Phys.\ Lett.\ B \textbf{597} (2004) 155 [hep-ph/0403016]; 
I.~de Medeiros Varzielas and G.~G.~Ross, 
Nucl.\ Phys.\ B \textbf{733} (2006) 31 [hep-ph/0507176]; 
G.~Altarelli and F.~Feruglio, 
Nucl.\ Phys.\ B \textbf{720} 64 (2005) [hep-ph/0504165]. 
X.~-G.~He, Y.~-Y.~Keum and R.~R.~Volkas, 
JHEP \textbf{0604} 039 (2006) [hep-ph/0601001]. 
R.~N.~Mohapatra and C.~C.~Nishi, 
Phys.\ Rev.\ D \textbf{86} 073007 (2012) [arXiv:1208.2875 [hep-ph]]. 
M.~-C.~Chen, J.~Huang, J.~-M.~O'Bryan, A.~M.~Wijangco and F.~Yu, 
JHEP \textbf{1302} 021 (2013) [arXiv:1210.6982 [hep-ph]]; H.~Ishimori and
E.~Ma, 
Phys.\ Rev.\ D \textbf{86}, 045030 (2012) [arXiv:1205.0075 [hep-ph]]. 
A.~E.~C\'{a}rcamo Hern\'{a}ndez, R.~Mart\'{\i}nez and F.~Ochoa, 
Phys.\ Rev.\ D \textbf{87} 075009 (2013) [arXiv:1302.1757 [hep-ph]]; 
A.~E.~C\'{a}rcamo Hern\'{a}ndez, I. d. M. Varzielas, S. G. Kovalenko, H.
P\"as and I. Schmidt, 
Phys.\ Rev.\ D \textbf{88}, 076014 (2013) [arXiv:1307.6499 [hep-ph]]; 
H.~Okada and K.~Yagyu, 
Phys.\ Rev.\ D \textbf{90}, 035019 (2014) [arXiv:1405.2368 [hep-ph]];
S.~F.~King, 
JHEP \textbf{1408} (2014) 130 [arXiv:1406.7005 [hep-ph]]; 
A.~E.~C\'{a}rcamo Hern\'{a}ndez, E.~C.~Mur and R.~Martinez, 
Phys.\ Rev.\ D \textbf{90}, 073001 (2014) [arXiv:1407.5217[hep-ph]]. 

\bibitem{Marzocca:2011dh} D.~Marzocca, S.~T.~Petcov, A.~Romanino and
M.~Spinrath, 
JHEP \textbf{1111}, 009 (2011) [arXiv:1108.0614 [hep-ph]].

\bibitem{Antusch:2013kna} S.~Antusch, C.~Gross, V.~Maurer and C.~Sluka, 
Nucl.\ Phys.\ B \textbf{877}, 772 (2013) [arXiv:1305.6612 [hep-ph]]. 

\bibitem{Chen:2013wba} M.~-C.~Chen, J.~Huang, K.~T.~Mahanthappa and
A.~M.~Wijangco, 
JHEP \textbf{1310}, 112 (2013) [arXiv:1307.7711 [hep-ph]].

\bibitem{King:2012in} S.~F.~King, C.~Luhn and A.~J.~Stuart, 
Nucl.\ Phys.\ B \textbf{867}, 203 (2013) [arXiv:1207.5741 [hep-ph]].

\bibitem{Meloni:2011fx} D.~Meloni, 
JHEP \textbf{1110}, 010 (2011) [arXiv:1107.0221 [hep-ph]].

\bibitem{BhupalDev:2012nm} P.~S.~Bhupal Dev, B.~Dutta, R.~N.~Mohapatra and
M.~Severson, 
Phys.\ Rev.\ D \textbf{86}, 035002 (2012) [arXiv:1202.4012 [hep-ph]].

\bibitem{Babu:2009nn} K.~S.~Babu and Y.~Meng, 
Phys.\ Rev.\ D \textbf{80}, 075003 (2009) [arXiv:0907.4231 [hep-ph]].

\bibitem{Babu:2011mv} K.~S.~Babu, K.~Kawashima and J.~Kubo, 
Phys.\ Rev.\ D \textbf{83}, 095008 (2011) [arXiv:1103.1664 [hep-ph]].

\bibitem{Gomez-Izquierdo:2013uaa} J.~C.~G\'{o}mez-Izquierdo,
F.~G.~'al.~Canales and M.~Mondrag\'{o}n, 
arXiv:1312.7385 [hep-ph].


\bibitem{Antusch:2010es} S.~Antusch, S.~F.~King and M.~Spinrath, 
Phys.\ Rev.\ D \textbf{83}, 013005 (2011) [arXiv:1005.0708 [hep-ph]]. 

\bibitem{Hagedorn:2010th} C.~Hagedorn, S.~F.~King and C.~Luhn, 
JHEP \textbf{1006}, 048 (2010) [arXiv:1003.4249 [hep-ph]].

\bibitem{Ishimori:2008fi} H.~Ishimori, Y.~Shimizu and M.~Tanimoto, 
Prog.\ Theor.\ Phys.\ \textbf{121}, 769 (2009) [arXiv:0812.5031 [hep-ph]].

\bibitem{Patel:2010hr} K.~M.~Patel, 
Phys.\ Lett.\ B \textbf{695}, 225 (2011)

\bibitem{Chen:2007} M.~-C.~Chen and K.~T.~Mahanthappa, 
Phys.\ Lett.\ B \textbf{652}, 34 (2007) [arXiv:0705.0714 [hep-ph]];

\bibitem{Chen:2003zv} M.~-C.~Chen and K.~T.~Mahanthappa, 
Int.\ J.\ Mod.\ Phys.\ A \textbf{18}, 5819 (2003) [hep-ph/0305088].

\bibitem{King:2013eh} S.~F.~King and C.~Luhn, 
Rept.\ Prog.\ Phys.\ \textbf{76}, 056201 (2013) [arXiv:1301.1340 [hep-ph]].


\bibitem{Campos:2014lla} M.~D.~Campos, A.~E.~C\'arcamo Hern\'andez,
S.~Kovalenko, I.~Schmidt and E.~Schumacher, 
Phys.\ Rev.\ D \textbf{90}, 016006 (2014) [arXiv:1403.2525 [hep-ph]].

\bibitem{Wang:2011ub} F.~Wang and Y.~-X.~Li, 
Eur.\ Phys.\ J.\ C \textbf{71}, 1803 (2011) [arXiv:1103.6017 [hep-ph]].

\bibitem{Georgi:1974sy} H.~Georgi and S.~L.~Glashow, 
Phys.\ Rev.\ Lett.\ \textbf{32}, 438 (1974).

\bibitem{Georgi:1979df} H.~Georgi and C.~Jarlskog, 
Phys.\ Lett.\ B \textbf{86}, 297 (1979).

\bibitem{Frampton:1979} P.~Frampton, S.~Nandi and J.~Scanio, Phys.\ Lett.\ B 
\textbf{85}, 225 (1979). 

\bibitem{Ellis:1979} J.~R.~Ellis and M.~K.~Gaillard, 
Phys.\ Lett.\ B \textbf{88}, 315 (1979).

\bibitem{Nandi:1980sd} S.~Nandi and K.~Tanaka, 
Phys.\ Lett.\ B \textbf{92}, 107 (1980).

\bibitem{Frampton:1980} P.~H.~Frampton, 
Phys.\ Lett.\ B \textbf{89}, 352 (1980).

\bibitem{Langacker:1980js} P.~Langacker, 
Phys.\ Rept.\ \textbf{72}, 185 (1981). 

\bibitem{Kalyniak:1982pt} P.~Kalyniak and J.~N.~Ng, 
Phys.\ Rev.\ D \textbf{26}, 890 (1982). 

\bibitem{Giveon:1991} A.~Giveon, L.~J.~Hall and U.~Sarid, 
Phys.\ Lett.\ B \textbf{271}, 138 (1991).


\bibitem{Dorsner:2007fy} I.~Dorsner and I.~Mocioiu, 
Nucl.\ Phys.\ B \textbf{796}, 123 (2008) [arXiv:0708.3332 [hep-ph]]. 


\bibitem{Dorsner:2006dj} I.~Dorsner and P.~Fileviez Perez, 
Phys.\ Lett.\ B \textbf{642}, 248 (2006) [hep-ph/0606062].


\bibitem{FileviezPerez:2007nh} P.~Fileviez P\`{e}rez, 
In *Karlsruhe 2007, SUSY 2007* 678-681 [arXiv:0710.1321 [hep-ph]]. 


\bibitem{Perez:2008ry} P.~Fileviez P\`{e}rez, H.~Iminniyaz and Germ\'{a}%
n~Rodrigo, 
Phys.\ Rev.\ D \textbf{78}, 015013 (2008) [arXiv:0803.4156 [hep-ph]]. 

\bibitem{Khalil:2013ixa} S.~Khalil and S.~Salem, 
Nucl.\ Phys.\ B \textbf{876}, 473 (2013) [arXiv:1304.3689 [hep-ph]]. 


\bibitem{Li:1973mq} L.~-F.~Li, 
Phys.\ Rev.\ D \textbf{9}, 1723 (1974). 















\bibitem{Bora:2012tx} K.~Bora, 
arXiv:1206.5909 [hep-ph]. 

\bibitem{Xing:2007fb} Z.~z.~Xing, H.~Zhang and S.~Zhou, 
Phys.\ Rev.\ D \textbf{77}, 113016 (2008) [arXiv:0712.1419 [hep-ph]]. 




\bibitem{Branco2010} G.~C.~Branco, D.~Emmanuel-Costa and C.~Simoes, 
Phys.\ Lett.\ B \textbf{690} (2010) 62 [arXiv:1001.5065 [hep-ph]];


\bibitem{CarcamoHernandez:2010im} A.~E.~C\'arcamo Hern\'andez and R.~Rahman, 
arXiv:1007.0447 [hep-ph]. 



\bibitem{Hernandez:2013hea} A.~E.~C\'arcamo Hern\'andez, R. Martinez and
F.~Ochoa, 
arXiv:1309.6567 [hep-ph]. 

\bibitem{Hernandez:2014vta} A.~E.~C\'arcamo Hern\'andez, R. Martinez and
Jorge Nisperuza, 
arXiv:1401.0937 [hep-ph]. 

\bibitem{Hernandez:2014a} A.~E.~C\'arcamo Hern\'andez and I.~de Medeiros
Varzielas, 
arXiv:1410.2481 [hep-ph].

\bibitem{Fritzsch} H.~Fritzsch, 
Phys.\ Lett.\ B \textbf{70}, 436 (1977), 
Phys.\ Lett.\ B \textbf{73}, 317 (1978), 
Nucl.\ Phys.\ B \textbf{155}, 189 (1979); 
H.~Fritzsch and J.~Planck, 
Phys.\ Lett.\ B \textbf{237}, 451 (1990). 

\bibitem{Xing} D.~s.~Du and Z.~z.~Xing, 
Phys.\ Rev.\ D \textbf{48}, 2349 (1993); 

\bibitem{FX} H.~Fritzsch and Z.~z.~Xing, 
Phys.\ Lett.\ B \textbf{353}, 114 (1995) [arXiv:hep-ph/9502297], 
Nucl.\ Phys.\ B \textbf{556}, 49 (1999) [arXiv:hep-ph/9904286], 
Phys.\ Lett.\ B \textbf{555}, 63 (2003) [arXiv:hep-ph/0212195]. 

\bibitem{Matsuda} H.~Nishiura, K.~Matsuda, T.~Kikuchi and T.~Fukuyama, 
Phys.\ Rev.\ D \textbf{65}, 097301 (2002) [arXiv:hep-ph/0202189]; 
K.~Matsuda and H.~Nishiura, 
Phys.\ Rev.\ D \textbf{69}, 053005 (2004) [arXiv:hep-ph/0309272]. 

\bibitem{Zhou} Y.~F.~Zhou, 
J.\ Phys.\ G \textbf{30}, 783 (2004) [arXiv:hep-ph/0307240]. 

\bibitem{Carcamo} A.~E.~C\'{a}rcamo Hern\'{a}ndez, R.~Martinez and
J.~A.~Rodriguez, 
Eur.\ Phys.\ J.\ C \textbf{50}, 935 (2007) [arXiv:hep-ph/0606190], 
AIP Conf.\ Proc.\ \textbf{1026} (2008) 272. 

\bibitem{Xing2010} Z.~-z.~Xing, D.~Yang and S.~Zhou, 
Phys.\ Lett.\ B \textbf{690}, 304 (2010) [arXiv:1004.4234 [hep-ph]];

\bibitem{Branco2012} 
G.~C.~Branco, H.~R.~C.~Ferreira, A.~G.~Hessler and J.~I.~Silva-Marcos, 
JHEP \textbf{1205}, 001 (2012) [arXiv:1101.5808 [hep-ph]];

\bibitem{Bhattacharyya:2012pi} G.~Bhattacharyya, I.~de Medeiros Varzielas
and P.~Leser, 
Phys.\ Rev.\ Lett.\ \textbf{109}, 241603 (2012) [arXiv:1210.0545 [hep-ph]]. 

\bibitem{King:2013hj} S.~F.~King, S.~Morisi, E.~Peinado and J.~W.~F.~Valle, 
Phys.\ Lett.\ B \textbf{724}, 68 (2013) [arXiv:1301.7065 [hep-ph]].

\bibitem{CarcamoHernandez:2012xy} A.~E.~C\'{a}rcamo Hern\'{a}ndez,
C.~O.~Dib, N.~Neill H and A.~R.~Zerwekh, 
JHEP \textbf{1202}, 132 (2012) [arXiv:1201.0878 [hep-ph]].


\bibitem{Vien:2014ica} V.~V.~Vien and H.~N.~Long, 
arXiv:1408.4333 [hep-ph].

%

\end{thebibliography}
\end{document}